\documentclass[preprint]{aastex61}
\usepackage{CJK}
\usepackage{amssymb}
\usepackage{bm}
\usepackage{color}
\usepackage{amsmath}
\usepackage{url}
\usepackage{listings}

\newcommand{\s}{\,{\rm s}}      \newcommand{\ps}{\,{\rm s}^{-1}}
    
\newcommand{\cm}{\,{\rm cm}}

        \newcommand{\K}{\,{\rm K}}

\newcommand{\cmmthree}{\,{\rm cm}^{-3}}

\newcommand{\dd}[1]{\mathrm{d}#1}

\newcommand{\xray}{{\rm X-ray}}

\begin{document}
\begin{CJK*}{UTF8}{bkai}


\title{Eigenvalue method for NEI unit in FLASH code}


\author[0000-0002-0747-0078]{Gao-Yuan Zhang (張高原)}
\affil{School of Space Science and Astronomy, Nanjing University, 163 Xianlin Avenue, Nanjing~210023, P.R.\ China}
\affil{Harvard-Smithsonian Center for Astrophysics, 60 Garden St., Cambridge, MA 02138 USA}
\author[0000-0003-3462-8886]{Adam Foster}
\author[0000-0003-4284-4167]{Randall Smith}
\affil{Harvard-Smithsonian Center for Astrophysics, 60 Garden St., Cambridge, MA 02138 USA}









\begin{abstract}
We describe an improved non-equilibrium ionization (NEI) method that we have developed as an optional module for the FLASH magnetohydrodynamic simulation code.  The method employs an eigenvalue approach rather than the earlier iterative ODE approach to solve the stiff differential equations involved in NEI calculations.  The new code also allows the atomic data to be easily updated from the AtomDB database. We compare both the updated atomic data and the methods separately.  The new atomic data are shown to make a significant difference in some circumstances, although the general trends remain the same.  Additionally, the new method also allows simultaneous calculation of the non-equilibrium radiative cooling, {which is not included in the original method}. The eigenvalue method improves the calculation efficiency overall with no loss of accuracy. We explore some common ways to present the non-equilibrium ionization state with a sample simulation, and find that using average ionic charge difference from the equilibrium tends to be the clearest method.
\end{abstract}




\section{Introduction} \label{sec:intro}

In astrophysical plasmas, the density can be so low that collisional
interaction
timescales can reach millions of years, leading to long delays between a
thermodynamic event and eventual ionization equilibrium. 
In a
collisional plasma, no ion population will be significantly impacted
by a density-weighted timescale (fluence) less than $n_e t\sim
10^7\cmmthree\s$ and equilibrium is not reached until $\ga 10^{12}\cmmthree\s$ 
\citep{Smith2010}. Therefore, the widely-used astrophysical 
assumption of collisional ionization
equilibrium (CIE)
fails in many scenarios, and the impact of non-equilibrium ionization (NEI)
must be considered.
Beyond purely theoretical analysis, 
observations of
supernova remnants \citep[SNR; e.\ g.,][]{Zhang2015}, 
intergalactic medium \citep[e.\ g.,][]{Yoshikawa2006},
and even possibly galaxy clusters \citep[e.\ g.,][]{Prokhovov2010}
have all shown
NEI signatures
in their
\xray\ spectra, including both ionizing and recombining features. 


Magnetohydrodynamic (MHD) simulations can reproduce 
many astrophysical scenarios, { likely to contain NEI plasmas, 
\citep[such as
SNR evolutions e.\ g.,][]{Zhou2011,Slavin2017}, but} the plasmas in the
simulations are often assumed to be CIE due to the additional overhead
and computations required by some NEI methods. If radiative cooling
from heavy elements is significant in the MHD simulation
or a detailed spectrum that can be compared to observations
is one of the { requirements} of the simulation,
the CIE assumption should be removed and an NEI calculation should
be done along with the simulation.



{ Most} current MHD simulation codes have not included 
the NEI calculation or the NEI unit is 
too slow for practical applications.
To investigate how 
an astrophysical thermodynamic event
influences the ionization states of the plasma, 
multi-dimensional MHD is required, which in turn requires
the NEI calculation to be as fast as possible.
With an improved simulation code and updated parameters,
the NEI unit { we show here} 
will be more convenient to use for research
involving collisional ionization in high energy astrophysics.
One option, if radiative losses and gas mixing are not significant,
is to approximate the effects after the initial MHD run. 
\citet{Shen2015} have developed a fast eigenvalue method 
that performs an NEI
post-process analysis of MHD simulations without integrating the NEI
into 
the simulation.
The FLASH code\footnote{\url{http://flash.uchicago.edu/site/}} \citep{Fryxell2000}
can perform the MHD simulation
with
an NEI unit in version 4.3
to
calculate the change in the ion population with
variations of the plasma's density and temperature \citep{Orlando2003a}.
But the NEI unit in the code uses
outdated ionization coefficients \citep{Summers1974}
and the current algorithm is inefficient.
{ It uses a Bader-Deuflhard semi-implicit ODE solver \citep{Bader1983} to solve a sparse system of stiff linear 
equations with
MA28\footnote{\url{http://www.hsl.rl.ac.uk/}}. It assumes that during a
hydro time step, the temperature and density remains unchanged and during
the hydro process the 
{advection} of different species is independent from the ionization or 
recombination. The accuracy of this method is mainly
determined by the accuracy 
of the atomic data (See a discussion in \S~\ref{sec:accuracy}). 
The radiative cooling is not considered in the original NEI unit which is
important for most of UV to X-ray emitting hot gas.}

In this paper, we describe
an eigenvalue method for 
NEI in MHD simulations (\S~\ref{sec:method}), with
comparisons to the existing method
for consistency (\S~\ref{sec:compare}).
We update the atomic data
in the original FLASH NEI code 
to use the updated rates \citep{Bryans2009} from
AtomDB\footnote{\url{http://www.atomdb.org/}}
\citep{Foster2013} for comparison. 
In the eigenvalue method,
a total radiative cooling from the plasma is also 
calculated (\S~\ref{sec:radiative}). 
Finally, a range of methods to measure
the NEI states are discussed with examples (\S~\ref{sec:discussion}).

\section{Eigenvalue method to solve the NEI problem}\label{sec:method}

\citet{Masai1984}, \citet{Hughes1985}, and \citet{Smith2010}
described an eigenvalue method to calculate NEI evolution, which we
briefly review below.
For a given atomic species, the fraction of the $i$th ionization state
can be derived from the differential equation set
\begin{equation}
    \frac{\partial{F_i}}{\partial{t}}=n_e\big\{\alpha_{i-1}(T)F_{i-1}-[\alpha_i(T)+R_{i-1}(T)]F_i+R_i(T)F_{i+1}\big\}
    \label{eq:1}
\end{equation}

where $n_e$ is the number density of electrons, $\alpha_i(T)$ is the
ionization rate from state $i$ to state $i+1$, and $R_i(T)$ is the
recombination rate from state $i+1$ to state $i$.
This equation includes only single ionizations and two-body
recombination. The method can be expanded, however, to include
multiple ionization and three-body recombination, if desired.
It can 
be written as
a matrix equation,
\begin{equation}
\frac{\partial{} \bm{F}}{\partial{\tau}}=\bm{A}\cdot\bm{F},
\end{equation}

where $\bm{F}$ is the ionization fraction vector for an atomic
species, $\dd{}\tau=n_e \dd{}t$ is the time scale,
and the tridiagonal matrix $\bm{A}$ contains 
the ionization rates and
recombination rates as shown in equation~\eqref{eq:1}.

In the numerical simulation, it can be assumed that the change of
temperature and density can be neglected when the time step is small enough.
For a given
temperature $T$, the equilibrium fraction, the eigenvalues, and
the eigenvectors of the matrix $\bm{A}$ 
are all constant, and thus
can be precalculated.
{ In the MHD run,
as long as the temperatures remain
within the allowable range, the
corresponding equilibrium fraction can
be found via interpolation by using the 
stored data tables. The eigenvalues and eigenvectors use the
nearest temperature in the data tables (the accuracy of this interpolation
is discussed
in \S~\ref{sec:accuracy}).}


The eigenvalue method can significantly accelerate the solution of 
differential equations, even though it requires more memory to store 
all the prepared
constant vectors than other ordinary differential equation (ODE) solvers.
{ Considering the equilibrium array, eigenvectors and eigenvalues matrix,
about 0.4 MB more memory is required than the original method for all the
twelve elements. It is worth
mentioning that both original and eigenvalue method need memories to store 
the ions fractions in each cell.}
{ To make it accurate enough, the original ODE solver needed to modify
the time step for solving the stiff equations. The
eigenvalue method, however, provides an exact solution to the stiff
equations
as long as the temperature and density can be assumed to be
constant within the time step.
The accuracy depends only on how accurate the eigenvalues and eigenvectors are computed.}

\section{The {\em neitest} model in FLASH code}\label{sec:compare}

The {\em NeiTest} problem in FLASH code (v4.3)
uses the NEI calculation unit to run a single test case.
The test assumes that
a plasma with a mass density of $2\times10^{-16}\,{\rm g}\cmmthree$ flows
with a constant uniform velocity of $3\times 10^5\cm\ps$
through a temperature jump going from $10^4\K$ to $10^6\K$. The plasma
is in ionization equilibrium before going through the jump in the
region at $T=10^4\K$.\footnote{FLASH code user guide. \url{http://flash.uchicago.edu/site/flashcode/user_support/flash4_ug_4p3/}}

\begin{figure}[htpb]
  \plottwo{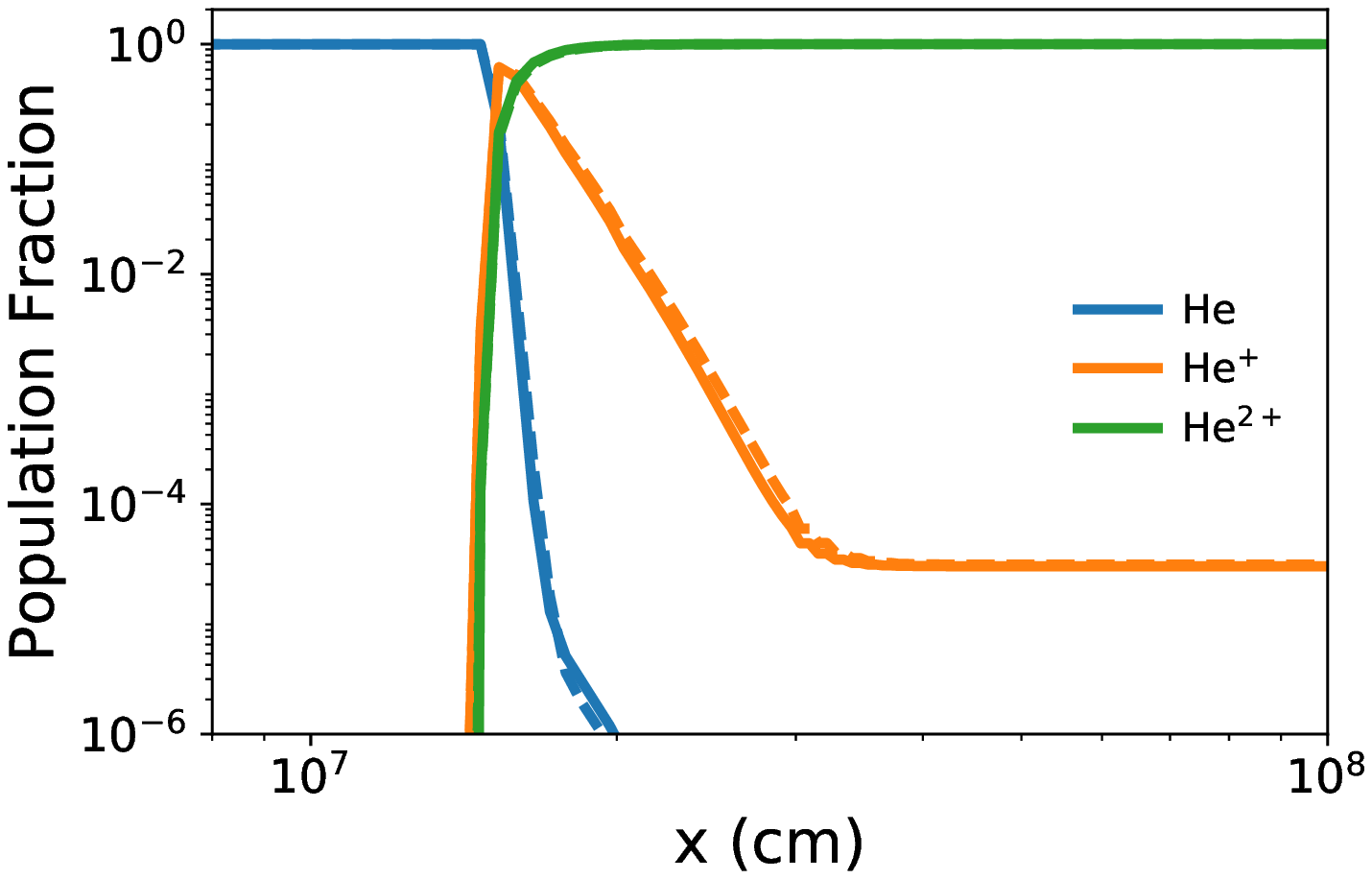}{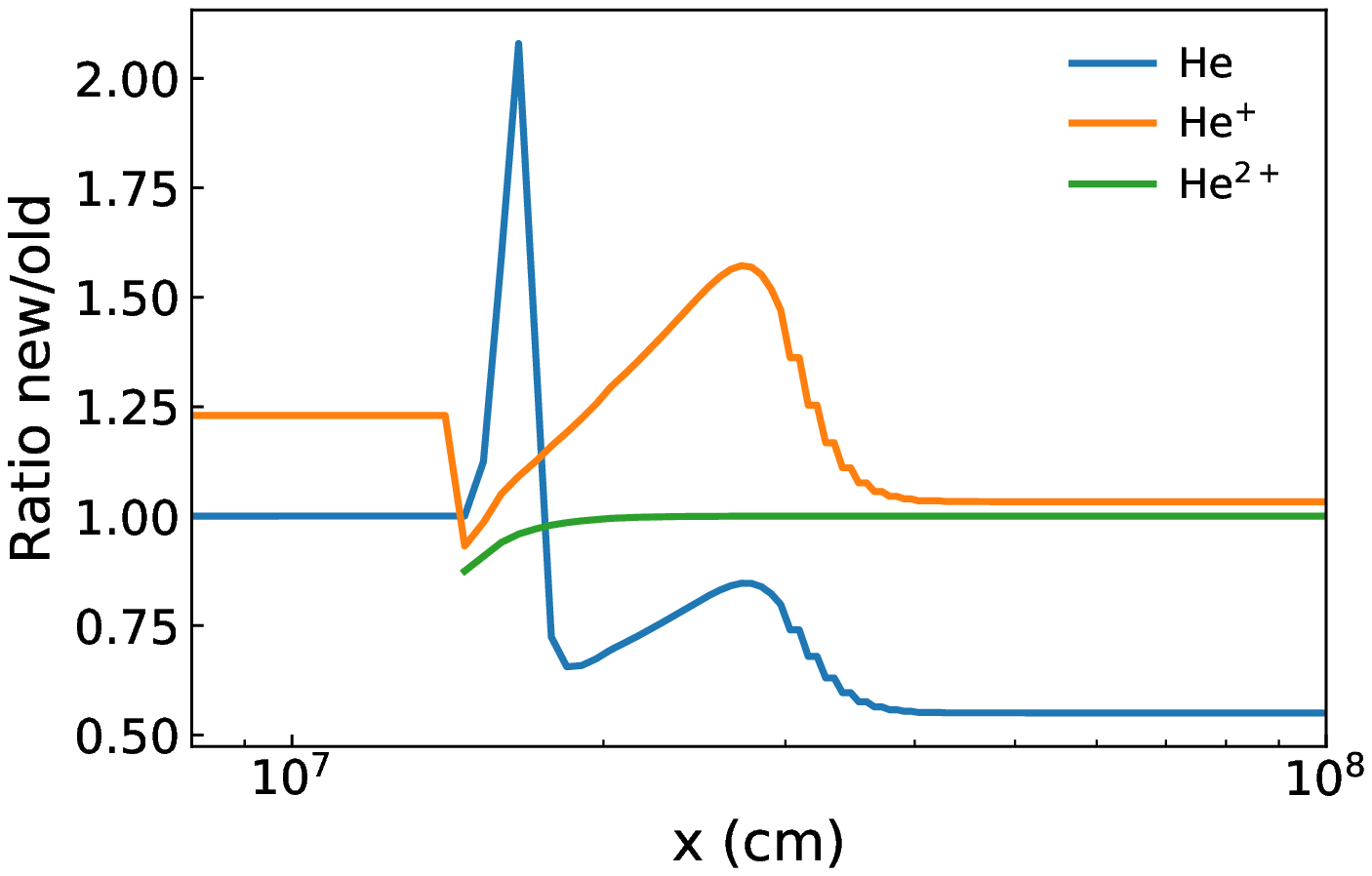}
  \caption{{ The population fractions evolution for He assuming a
    stationary flow through a temperature jump with the original ODE
    solver to compare the different
    ionization rate coefficients. Left panel: the result
  from the original ionization coefficients (solid lines) and the result
  from the new ionization coefficients (dashed lines). Right panel:
  the ratio of the results from the new ionization coefficients
  to the original one. 
  \label{f:he}}}
\end{figure}

We set up the simulation with the default parameters in one-dimension,
running with 8 cpu cores using an Intel Fortran 2017 compiler.
The results are recorded 
with an interval time of 50$\s$.
The code stops when the time reaches 1000$\s$.
The python-based yt-project\footnote{\url{http://yt-project.org/}} \citep{Turk2011} is
used for the analysis of the results.
{ The flow from lower temperature region to higher region can show the
ionization along the time as the velocity is constant. After the ``shock 
front'' goes through the simulation regime, the
ionization state at each point will become stable, which is not necessarily 
equilibrium. It is reasonable to use such a stable state to do post-process
analysis. Considering the constant velocity and the largest spatial distance,
it has become stable at $500\s$, and the result files 
can be used for the post process analysis.}
{ See
Fig.~\ref{f:he} for the results for He ions as an example.
}

\subsection{Impact of the updated atomic data}\label{sec:parameter}

{ As mentioned in \S~\ref{sec:intro}, in the original NEI unit,
a semi-implicit ODE solver to
solve stiff equations
is used for the
evolution of ionization fractions.}
The table file ``summers\_den\_1e8.rates'' contains
ionization and recombination rates for
He, C, N, O, Ne, Mg, Si, S, Ar, Ca, Fe, Ni taken from \citet{Summers1974}.
We substituted this file with one containing updated rates 
assembled by \citet{Bryans2009}.
All other
parameters and settings are remained
the same to allow comparison.

\begin{figure}[htpb]
\centering
 \plottwo{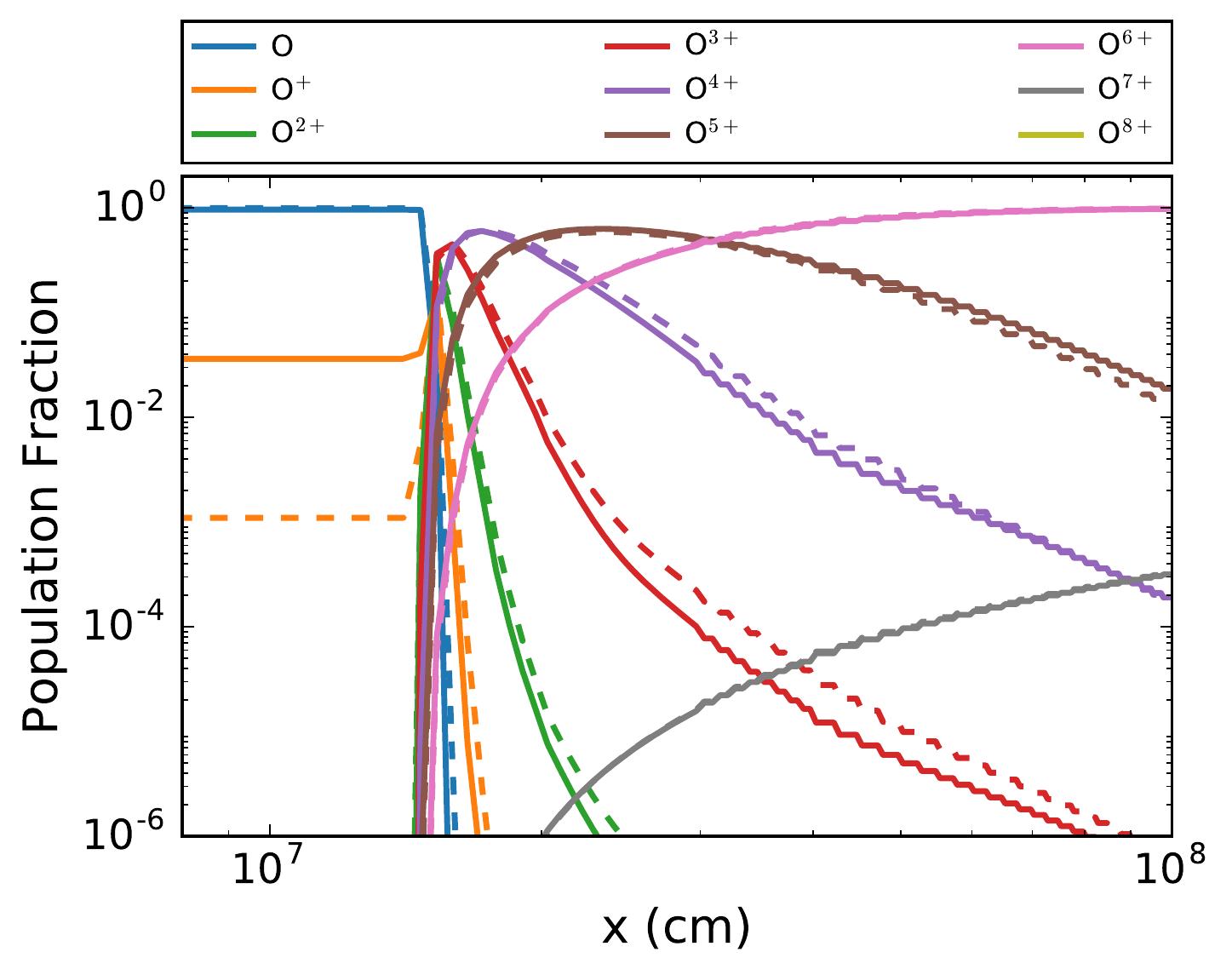}{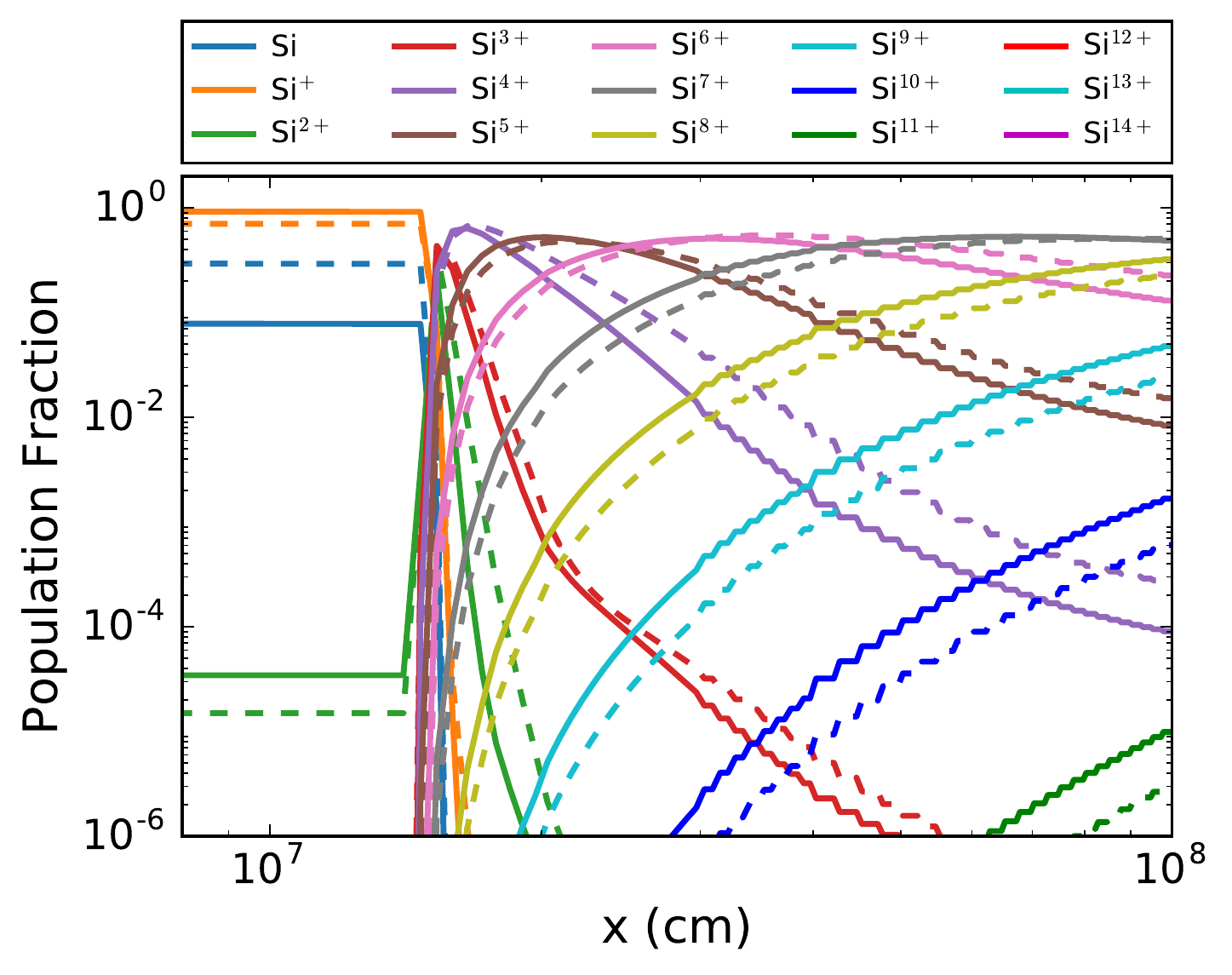}
 \plottwo{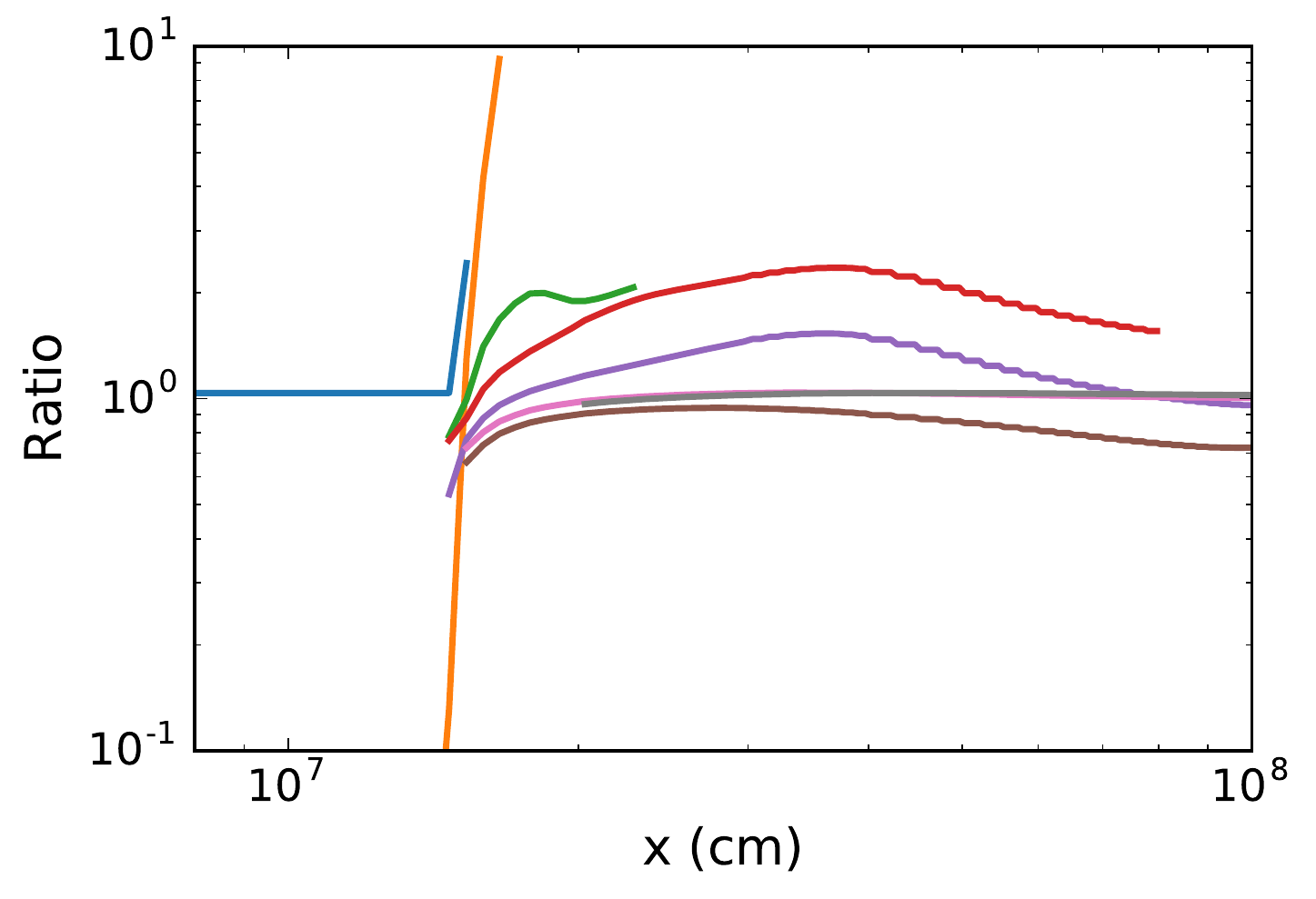}{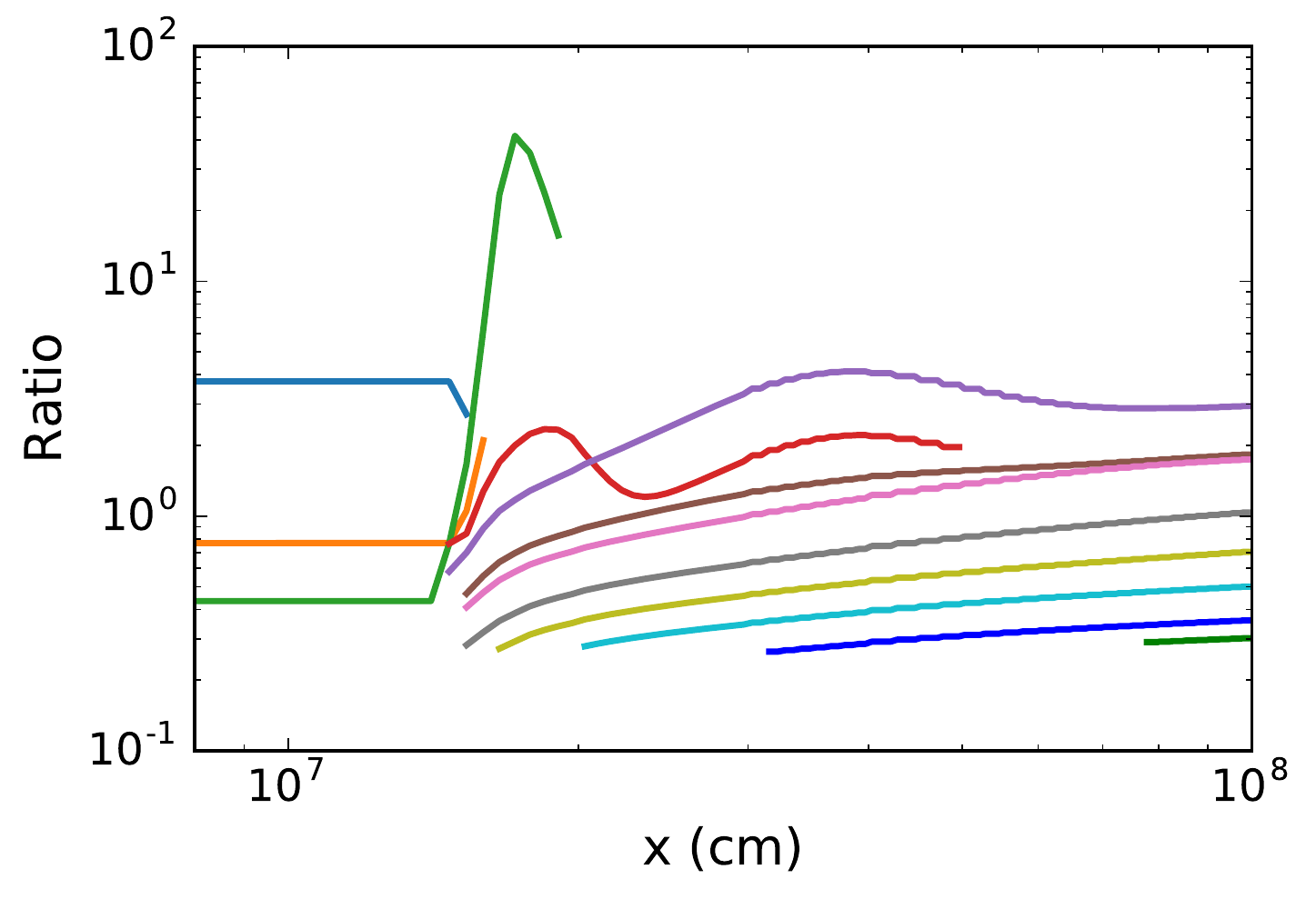}
 \caption{The population fractions ratios between the new atomic data and
    the original data for O (left) and Si (right) as examples. { In the 
    top panel,} the result
  from the original ionization coefficients is displayed in solid lines
  and the result
  from the new ionization coefficients in dashed lines.
   Bottom panel is the ratio of the new 
  to the original method coded in the same color as the top panel.
  It is not shown when the value for either method is less than 
    $1\times10^{-6}$.\label{f:ratios}}
\end{figure}

We have checked the differences for all the ions in the ``summers\_den\_1e8.rates''
file.
Here, we only present the
ratio between the updated one and the original one of
He, O, and Si to show the differences 
(ignoring values below 1$\times 10^{-6}$).
The results show that the new ionization coefficients can cause
significant differences in the ionization test module (see
Fig.~\ref{f:he} and Fig.~\ref{f:ratios}).
As expected, the overall trends
for both the old and new ionization rate are the same.
However, from the ratio
figures, the ion fraction shows significant differences from the initial state to
the final state, especially
for heavier elements.
This difference should be considered in the
simulation or analysis of ionization in a plasma.
In Fig.~\ref{f:ratios}, we can see a ``step'' shape in lines, it is because
FLASH code uses an adaptive mesh refinement (AMR) grid and the spatial 
resolution is not the same all over the simulation regime. It is sparser
for a constant density on the right side.

%

\subsection{Test the eigenvalue method}

To compare the eigenvalue method with the original method, we also
used the {\em NeiTest} simulation. The old ODE method code was run with
the updated parameter table file as described in \S~\ref{sec:parameter},
ensuring the codes are using 
the same atomic data.

\begin{figure}[htpb]
  \plottwo{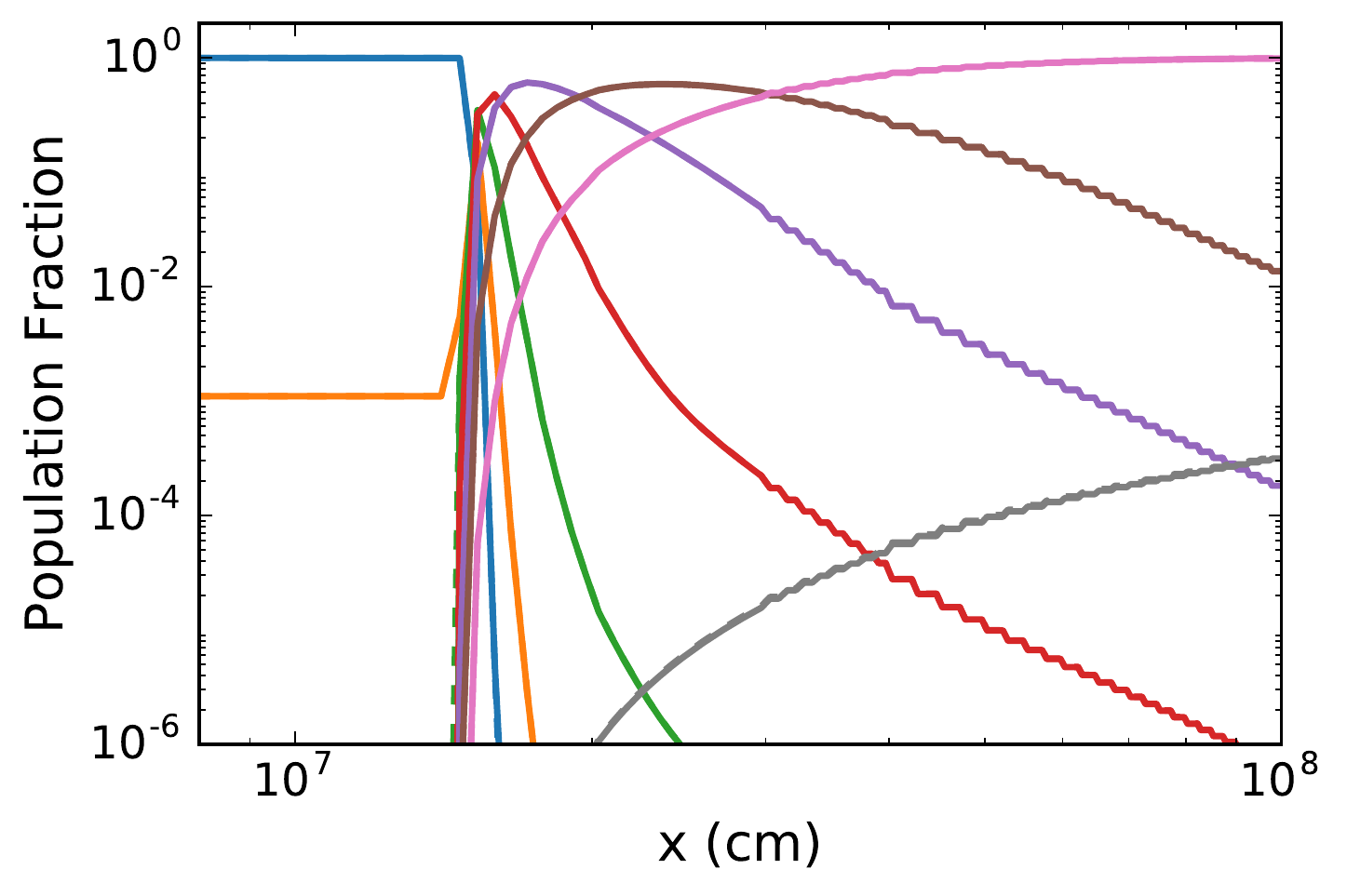}{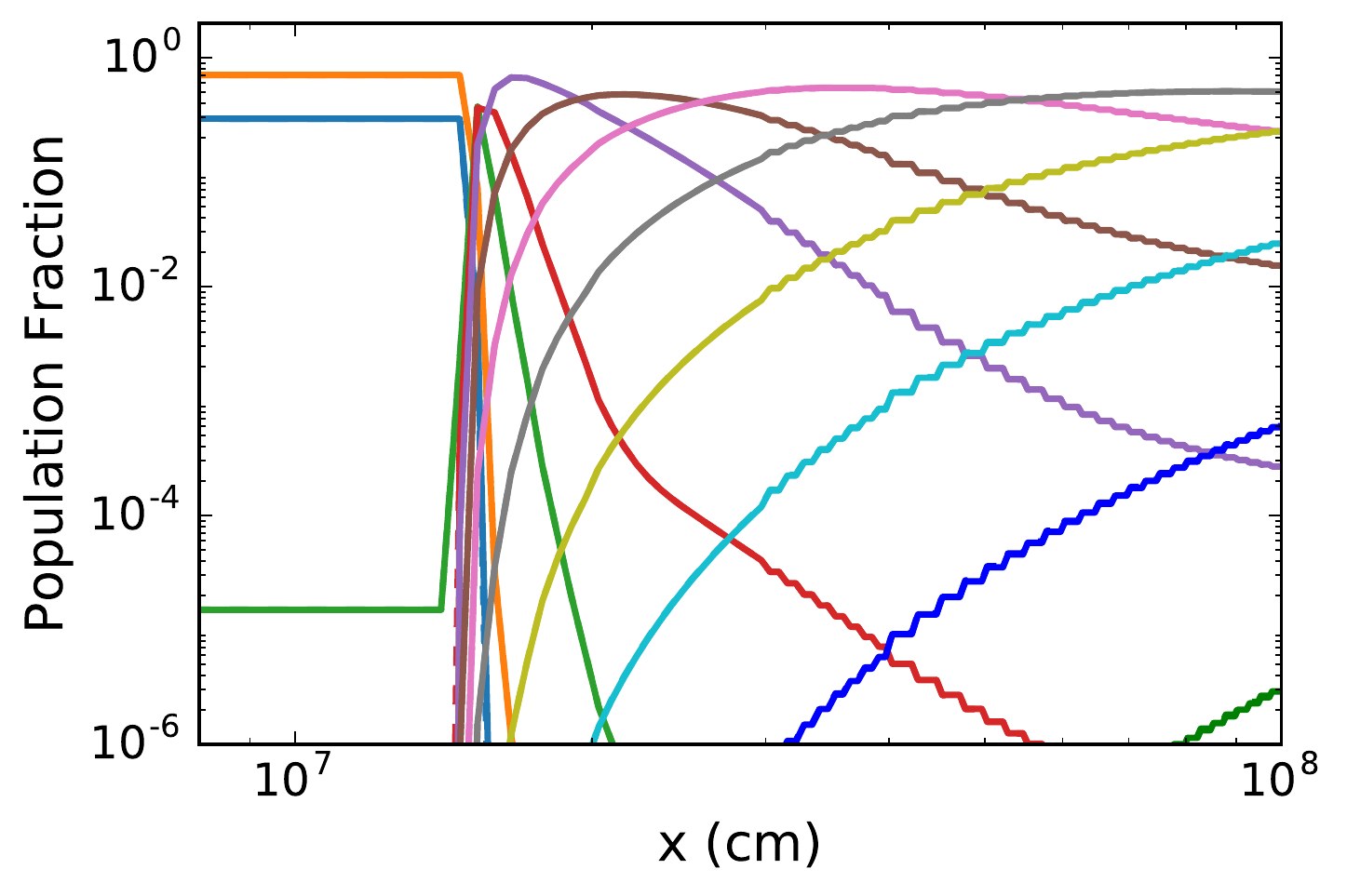}
  \plottwo{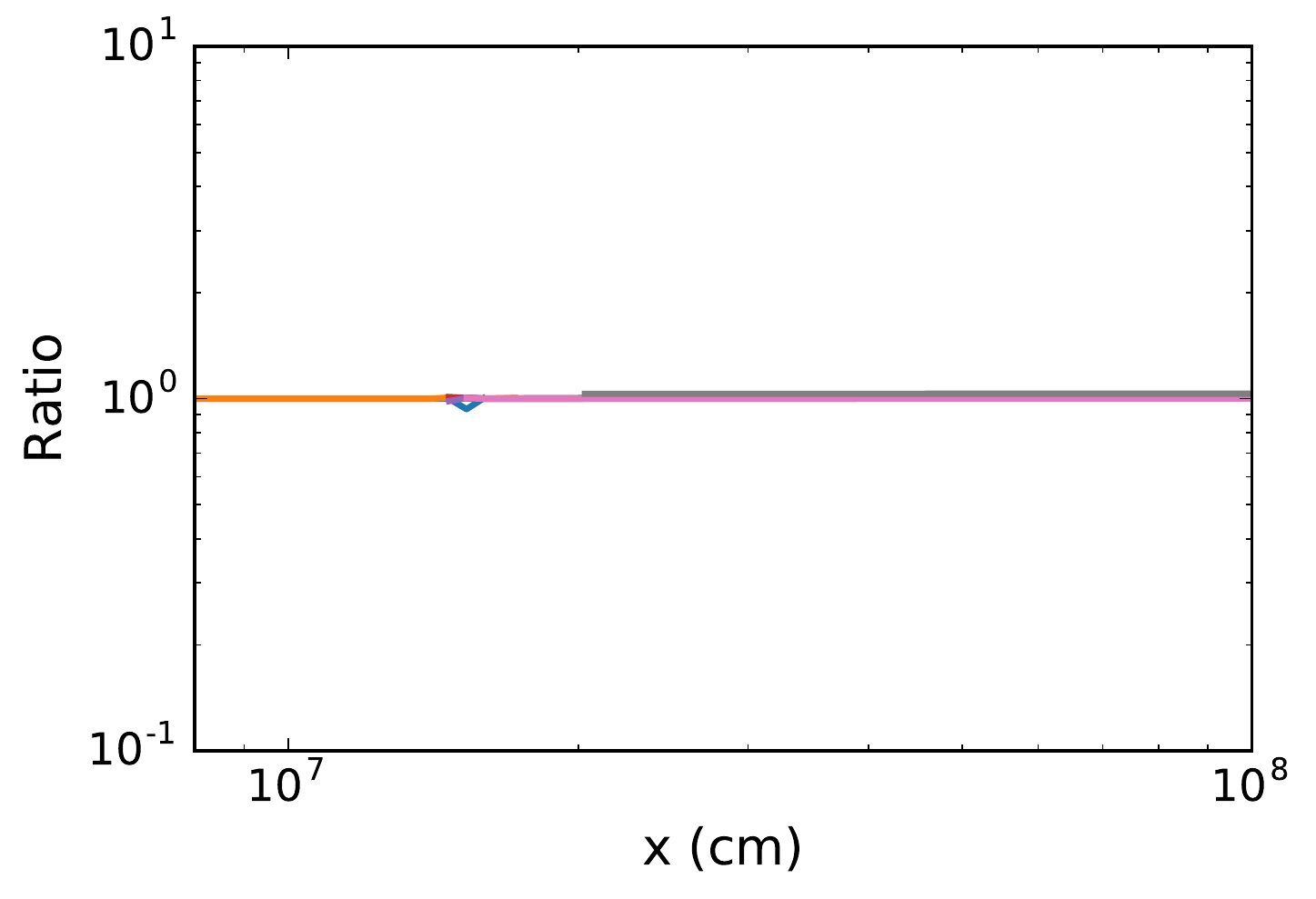}{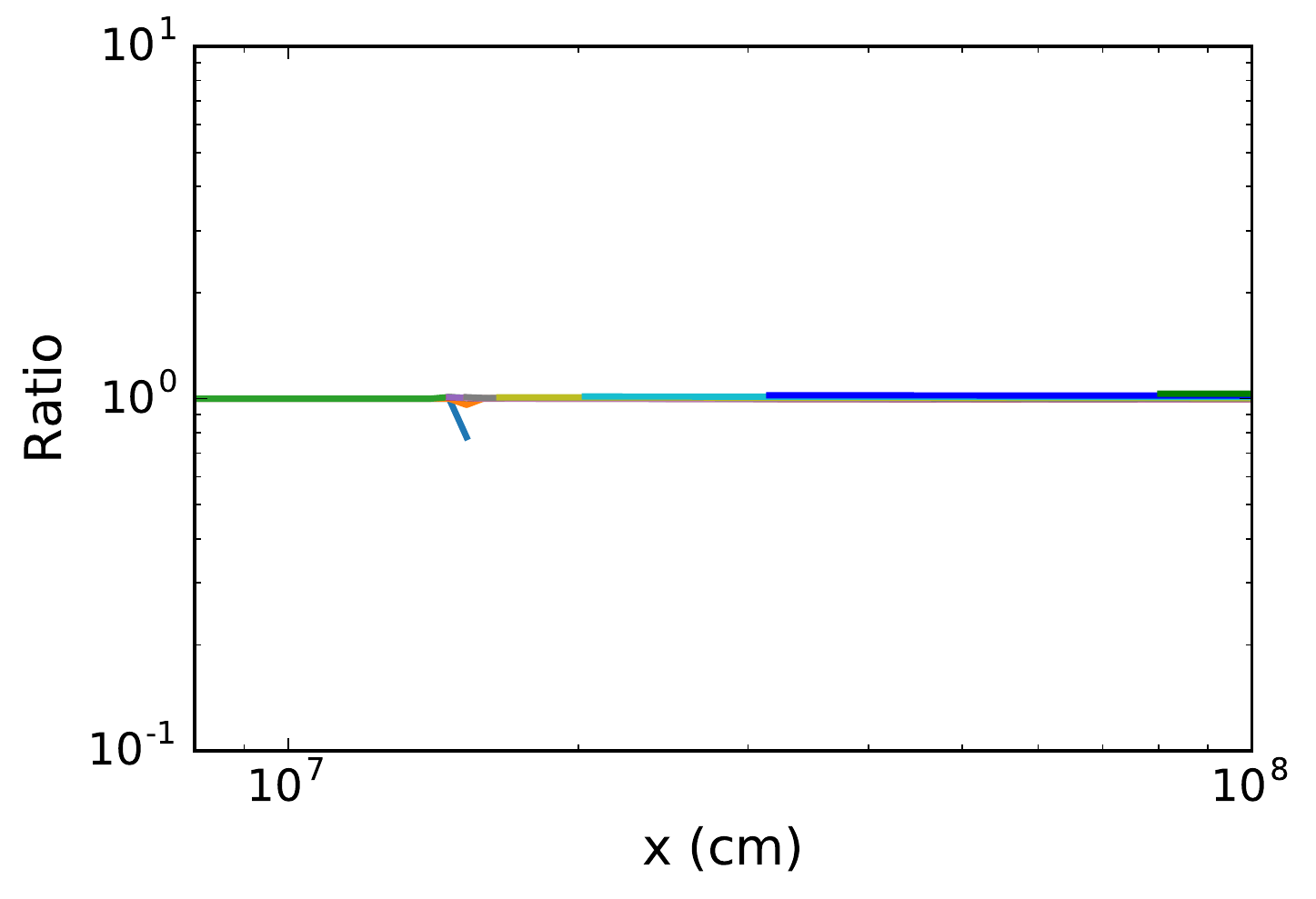}
  \caption{The comparison between the eigenvalue method and the
    standard ODE (MA28) method for O (left) and Si (right). The solid lines
    show the results from eigenvalue method; the dashed lines are from
    the original method (with the updated parameter table file). 
    Bottom panel is the ratio of the new 
    to the original method coded in the same color as the top panel.
    It is not shown when the value for either method is less than 
    $1\times10^{-6}$. All colors are coded the same with Fig.~\ref{f:ratios}
    \label{f:comparison}}
\end{figure}

In Fig.~\ref{f:comparison}, { the O and Si ions are shown
as examples to compare
the simulation results for the eigenvalue method and the original
method used in the old NEI code.} Because they use the same
atomic data, the two methods are consistent with each
other for the equilibrium state before and after the
{ temperature jump. Beyond the jump}, 
the values at very low
populations differ slightly, 
mainly because these two methods 
use different thresholds for the lower limits on the fraction values.
{ In the original method, 
a threshold of mass fraction is 1$\times 10^{-30}$;
while in the eigenvalue method, a threshold of about $1\times 10^{-10}$ 
of ion fraction is used 
for the eigenvalues and eigenvectors. Therefore, the states
of ions have a very small difference between each other. 
Considering the accuracy of the 
methods and the atomic data (See \S~\ref{sec:accuracy}), it is negligible.}
These ion populations with low values do not have much influence on the
final results of the major population at that temperature.

\begin{figure}[htpb]
  \plottwo{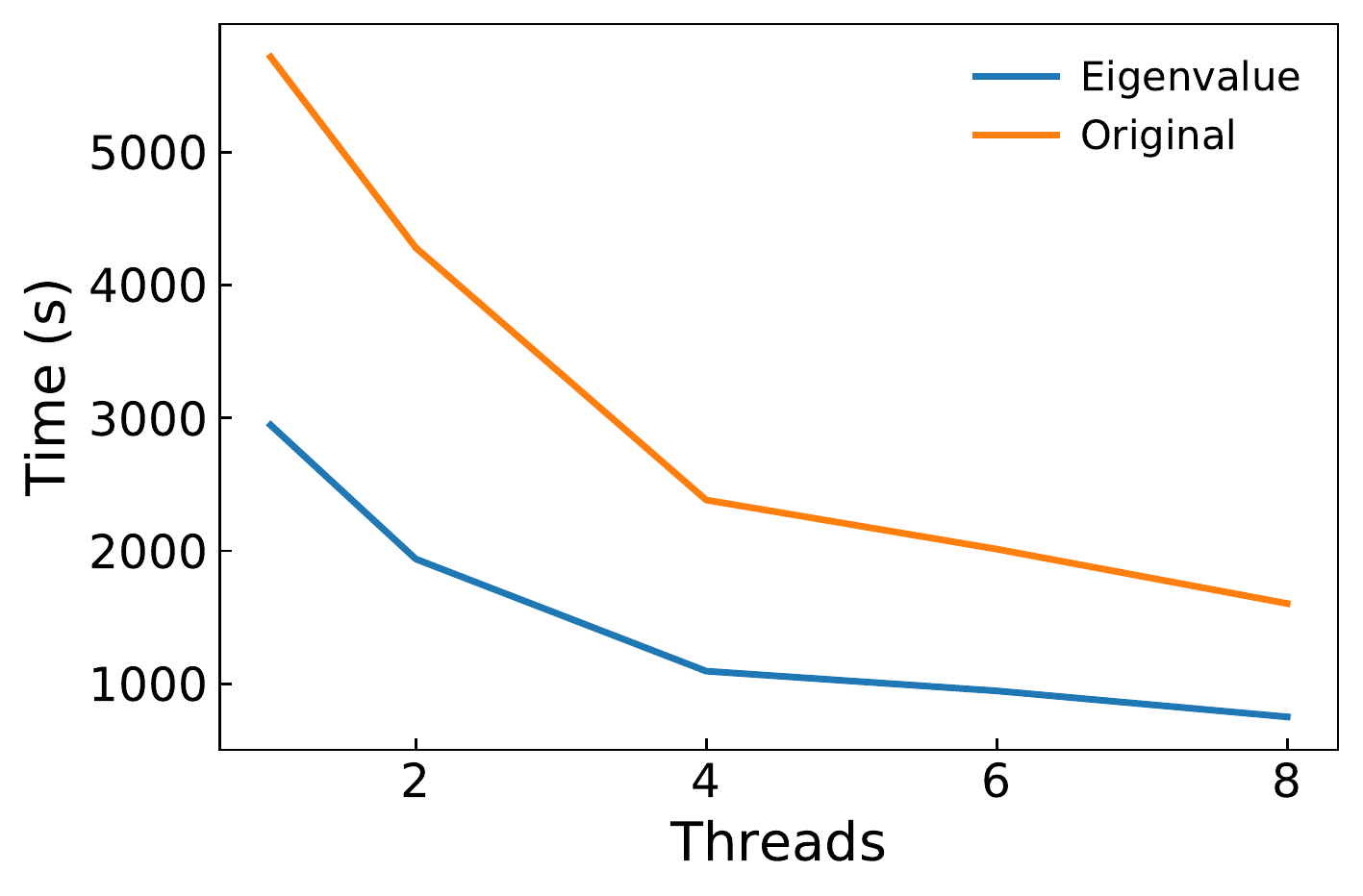}{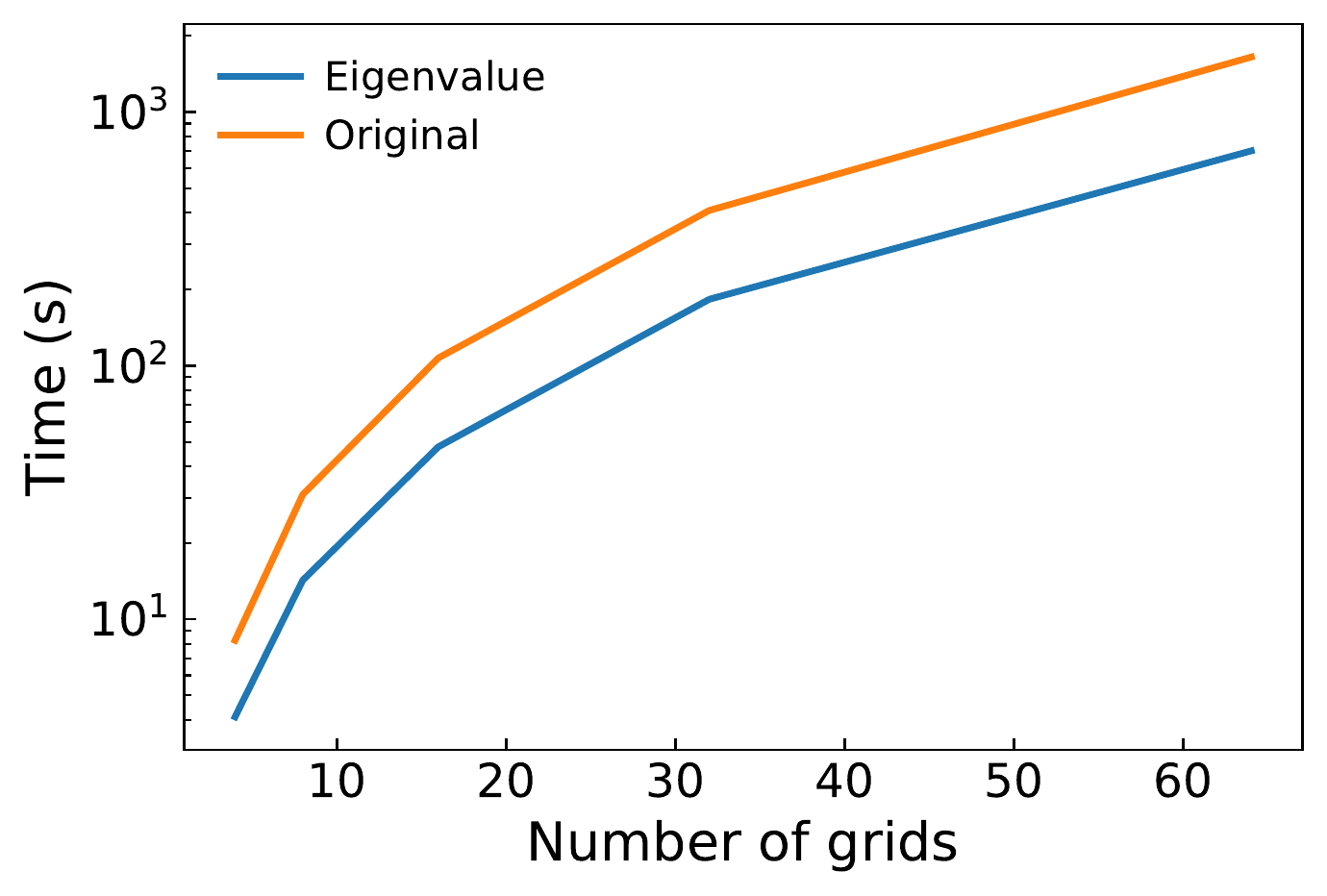}
  \caption{Consumed time (``evolution'') for different number of calculation 
  threads (Left) and for different simulation grids (Right).\label{f:cpu}}
\end{figure}

To compare the performance of the eigenvalue method and the original method,
{ two models were performed, ``NeiTest'' which is also the model used for 
the previous tests and a 2D parallel plane shock.
The eigenvalue method needs to load the matrices to calculate the ionization
in each cell, and this
initialization process consumes more time than the 
original method. Table~\ref{tab:time} has already separated the presentation
of the ``initialization'' and ``evolution'' calculation time. For a larger
simulation, ``initialization'' time can be neglected comparing to the whole
simulation time. So we compare the ``evolution'' time of both methods here. 
}
{ By using a different number of threads for the same 
calculation, both methods
provide an acceleration as more threads are used (See left panel of
Fig.~\ref{f:cpu}). The ratio between them remains similar for different
number of threads. Because the FLASH code can use an AMR grid, the AMR refinement or the number of blocks
of the simulation is another factor for the performance. With different 
size of fixed grids,
the performance of both methods is compared in the right panel of Fig.~\ref{f:cpu} (four threads are used). 
Table~\ref{tab:time} shows the results when the refinement level of AMR is free
to change in
a range from 1 to 6 for the two models. The plane shock model may change the block
number as the shock front moving through the simulation box.
From all the above tests, the eigenvalue method can be much more efficient
than the original one (more than a factor of two.)}

\begin{deluxetable}{ccc|ccc|ccc}
\tabletypesize{\footnotesize}
  \tablecaption{NEI calculation performance\label{tab:time}}
  \tablehead{
    \colhead{id} & \colhead{Model} & \colhead{Speedup\tablenotemark{$\dagger$}} 
    & \multicolumn{3}{c}{Eigenvalue
      Method} & \multicolumn{3}{c}{ODE (M28)} \\
    \cline{4-9}
      & & & \colhead{Total
      time (s)} & \colhead{Initialization (s)} &
    \multicolumn{1}{c|}{Evolution (s)} & \colhead{Total time (s)} &
    \colhead{Initialization (s)} & \colhead{Evolution (s)}
  }
  \startdata
  1    & NeiTest   & 1.67 &  104.105 & 2.942 & 101.158 &  172.575 & 0.113 & 172.458 \\
  2    & NeiTest   & 1.67 &  103.192 & 2.992 & 100.197 &  172.516 & 0.146 & 172.366 \\
  3    & NeiTest   & 1.67 &  103.309 & 2.996 & 100.309 &  172.486 & 0.140 & 172.342 \\
  4    & NeiTest   & 1.68 &  99.828  & 2.859 & 96.965  &  168.298 & 0.140 & 168.154 \\
  5    & Sod shock & 2.51 &  131.129 & 6.462 & 124.634 &  329.777 & 1.470 & 328.777 \\
  6    & Sod shock & 2.49 &  130.873 & 6.544 & 124.414 &  326.245 & 1.469 & 324.740 \\
  7    & Sod shock & 2.52 &  130.911 & 6.544 & 124.332 &  330.040 & 1.455 & 328.553 \\
  8    & Sod shock & 2.55 &  131.434 & 6.533 & 124.869 &  335.459 & 1.518 & 333.894 \\
  \enddata
\tablenotetext{\dagger}{Ratio of the total time between
M28 and Eigenvalue method}
\end{deluxetable}

\section{Radiative cooling model}\label{sec:radiative}

A radiative energy loss term
has also been added to the NEI code, with a variable
to switch it on and off. When it is switched on, the radiant energy
density in unit time of each ion species (${s}_{rad}^{(z,i)}(T)$ in unit of
$\rm{erg}\cdot\rm{cm}^3\cdot s^{-1}$) is retrieved
from a database table based on the temperature. Here, $z$ and $i$
imply the atomic number and the ionization states respectively.
Including the number densities and the volume, the energy loss rate is
written as $\dot{u}_{rad}^{(z,i)}=\int s_{rad}^{(z,i)}\cdot n_e\cdot n_{z,i}\cdot
dV$.
Therefore,
the total decrease on energy is the summation of every species
$\Delta u=\sum\limits_z\sum\limits_i\dot{u}_{rad}^{(z,i)}\cdot\Delta t$. 
MHD codes such as FLASH are developed with an adaptive grid
which makes it difficult to store the total energy. Instead, an
energy variable in unit mass, $u_m=\Delta u/(\int \rho dV)$, can be used, 
which is also consistent with other energy variables in FLASH code. The mass
density $\rho_i$ is connected to the number density of the species $n_i$ by
$\rho_i=(\mu/N_A) n_i$, where $\mu$ is the atomic weight of the species, 
$N_A$ is the
Avogadro's number. If it is for the whole gas, $\mu$ is the average
atomic weight.
The internal energy will subtract the energy loss at every step,
and the temperature
changes with the internal energy; the kinetic energy
will then be impacted indirectly. 

{ From the exponential results of eigenvalue method,
$n_{z,i}$ can be calculated precisely during a 
hydro time step. With the integration, an exponential term ($(e^{\lambda_i n_e\Delta t}-1)/\lambda_i$) can be used instead of $\Delta t$. 
Currently we use the first-order approximation for a faster performance, 
and more importantly to make it an independent 
code unit that does not depend on the eigenvalue ionization calculation.
It is a good estimation as long as 
the absolute value of $\lambda_i n_e\Delta t$ is small enough.}

During the time step $\Delta t$ that is determined by the MHD evolution
code (Courant-Friedrichs-Lewy; CFL condition),
it is assumed that the temperature ($T$),
the electron number density ($n_e$), and
the fraction of each species that affects the ions' number density
($n_{z,i}$) remain unchanged. When using the eigenvalue method, the
calculation of ionization does not require a very small time
step. However, the radiative cooling depends on and affects the
temperature, density, and the variation of the ion
fractions, which demand a time step small enough to make sure the
energy loss can be assumed negligible.
A new (smaller) time step will be required if the radiative loss
exceeds a threshold.
The new limit for time step is 
\begin{equation}
\Delta t_{rad}=c_{rad}\frac{E_{int}}{|\Delta u|}
\Delta t
\end{equation}
where $\Delta t$ is the current time step, 
$E_{int}$ is the internal energy, and
$c_{rad}$ is a constant to constrain the new
time step. { To be compatible with the CFL 
condition and make MHD simulation stable, $c_{rad}$ must be less than one.
Although it can be adjusted to balance the requirement of 
accuracy and efficiency, we recommend a $c_{rad}$ less than 
1$\times 10^{-2}$ to make sure that the first-order approximation and the assumption of an unchanged temperature 
within a time step are valid.}
The time step used by the simulation will
chose the smallest value among the time
steps determined by all physical conditions.

To test the radiative cooling model, { a static simulation with
a constant volume was performed
with the initial temperature $T=1\times10^6\K$, constant density
$n_H=0.1\cmmthree$, and $c_{rad}=1\times 10^{-2}$. Fig.~\ref{f:radi} shows that the results from 
the FLASH code and the same calculation of radiative cooling
with python module (pyAtomDB\footnote{\url{http://atomdb.readthedocs.io/en/master/}}).
In the python calculation, the temperatures are directly adopted from FLASH code,
but the ionization and emission energy rate are calculated from pyAtomDB.
\begin{figure}
\plottwo{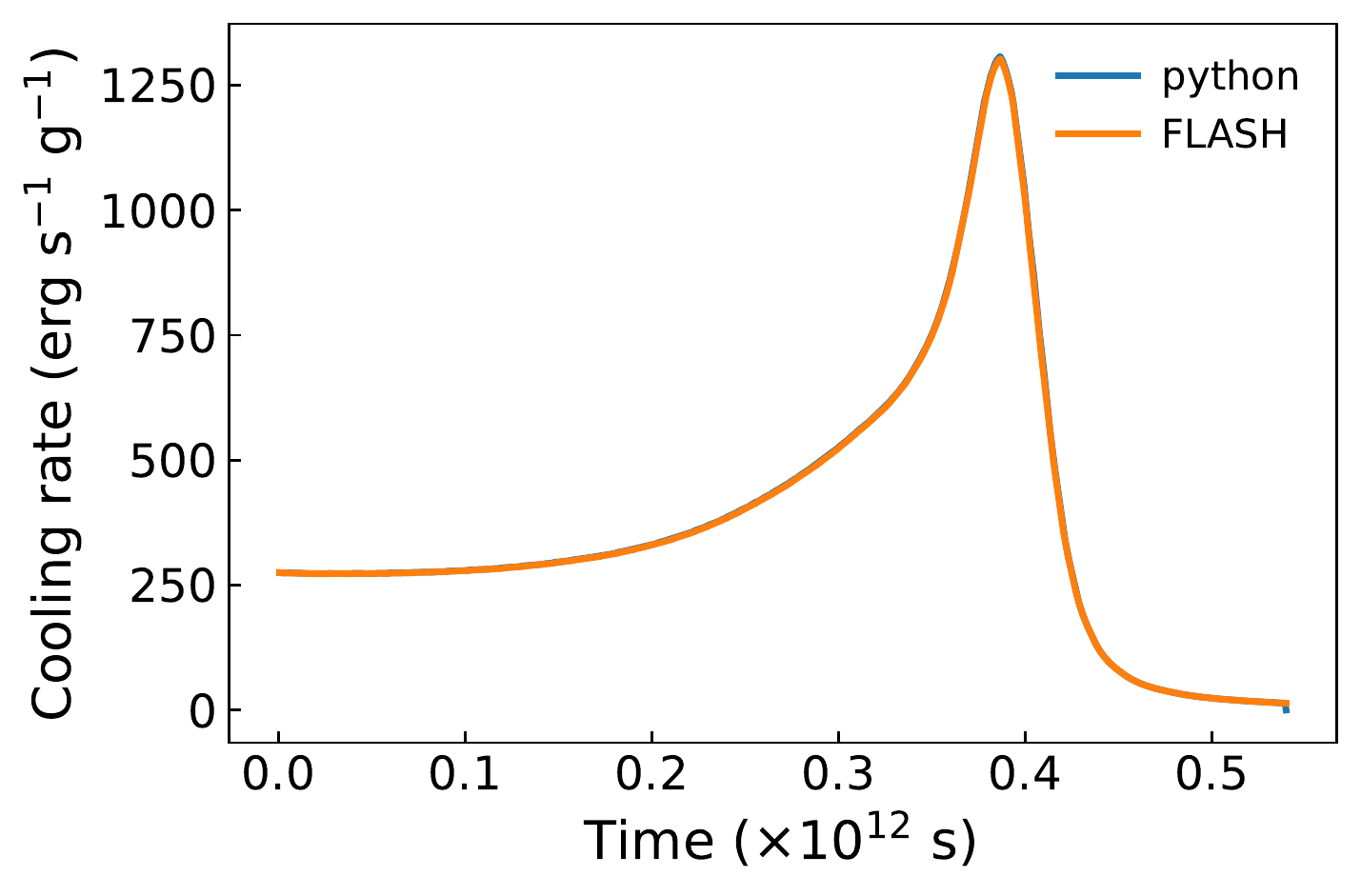}{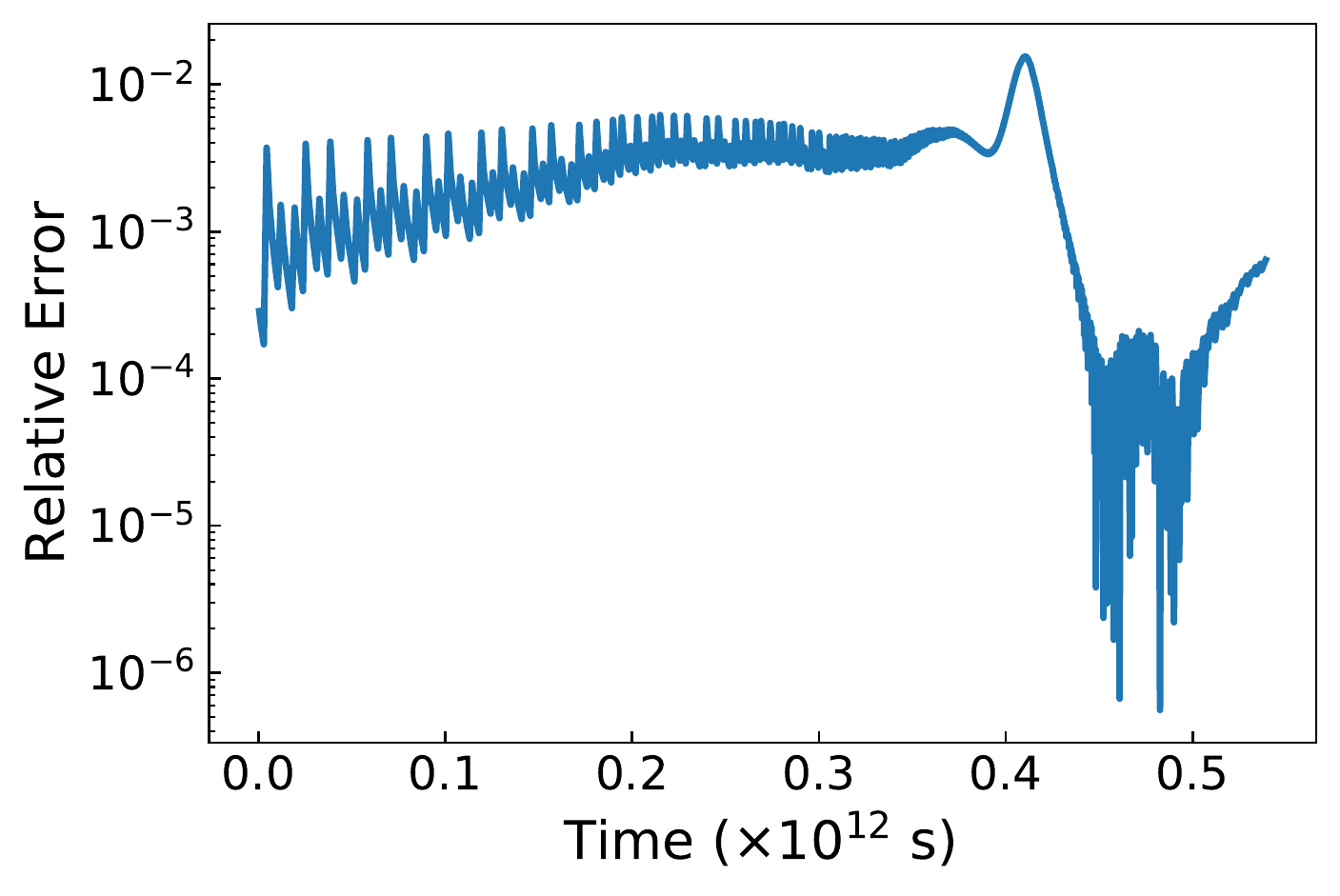}
\caption{Cooling rate changes with time in a constant volume cell 
when the initial temperature is $1\times 10^6\K$ (Left),
and the relative error correspondingly (Right).\label{f:radi}}
\end{figure}
The calculations are consistent with each other. When the cooling rate increases 
abruptly at about 4$\times10^{11}\s$, temperature falls
to 1$\times10^4\K$ and the time
step becomes smaller.}
{ We also perform several calculations with different initial temperatures. They 
show a similar or even smaller relative error.}
{ Because the NEI in the python scripts
is calculated independently from the FLASH 
MHD simulation, we can also conclude that the NEI error does not accumulate.}
%
The range of
the temperature for the radiative cooling is
($1\times10^4\K$--$1\times10^9\K$). The cooling process will not fail but print
warnings if the temperature gets out of this range, because the data
we use for the extended range of temperature are less accurate.
As we focus on the \xray\ emission here, the temperature range should
be kept within this range.


%

\section{Discussion}\label{sec:discussion}

\subsection{Accuracy of eigenvalue method}\label{sec:accuracy}
{
With the assumption that in a time step the temperature and 
density remain unchanged, the eigenvalue method is accurate no matter 
how large the time step is. However, there are some other sources of 
errors, such as the interpolation in temperature regime and the precision 
of the stored data. 
The precision can be made smaller 
by making sure the significant figures in the data are sufficient.

The interpolation is the main source of the inaccuracy here. 
From temperature $1\times 10^{4}\K$ to $1\times 10^{9}\K$,
matrices of eigenvalues and eigenvectors for 5000 temperature 
nodes are generated. At a temperature node, the eigenvalue method with
the matrices is accurate without error from interpolation.
We also calculate the ionization cases at the 5000 temperature nodes
with sparser matrices (1251 different temperature
nodes) that is used in the new NEI unit of FLASH code.
By comparing to the accurate results, the maximum of relative error 
($|f_{1521}-f_{5000}|/f_{5000}$) 
is less than 5\% 
throughout the temperature range. 
This was deemed acceptable considering that
the error from atomic data \citep{Dere2007} is about 10-15\%, which is 
also true for the original ODE method. 
Therefore, currently the error for the NEI solution is mainly from
the inaccuracy of the atomic data. 
A denser set of temperature matrices can be used
to increase the accuracy when more accurate atomic data are available.
We also tested the ionization cases with hundreds of steps and a big step
covering the total time of small steps. The interpolation error does not
accumulate.}

\subsection{Effect of the ionization of He}
{
He is not a minority species. The ionization of He can affect 
both the electron density and the temperature even during a time step.
When the temperature has been
over $\sim 1\times10^{5}\K$ for more than $n_e t\sim 10^{12}\cm^{-2}\s$,
He will be essentially
fully ionized and it will not change electron density and temperature.
However, when the temperature is under $\sim 1\times10^{5}\K$, the ionization
or recombination
of He can change the electron density by at most 20\%. Ionization of He will 
also carry a fraction of energy from the hydrodynamic process, 
leading to a change in 
the temperature. The calculations
relating to this temperature range or lower should be used carefully.

These effects will be added as
a change of the electron density and 
a cooling procedure for ionization to the FLASH code in our future work. 
The same problem is even more severe with H that is assumed fully ionized 
in our simulations.
}
\subsection{Methods to measure the ionization state}

{ As a result from the stiff ODE,} the difference between
the NEI and equilibrium decreases exponentially, as
the eigenvalues are negative. For an NEI plasma, we are generally
interested in
what kind of state (ionizing or recombining) the plasma is in, as well
as the extent to which this state 
deviates from the equilibrium.
We use results from
an SNR simulation (taken from work in preparation) to display the effect
of the different methods (Fig.~\ref{f:discussion}). The simulation is set
to be a SNR explosion going through some spherical dense clouds \citep[in 2D;][]{Slavin2017}. 
The ionization and recombination appear in the shock front and around the
clouds.

\subsubsection{Ratio between two different ionization states}
\label{sec:lineratio}
In many observations, the line emission ratios can be used to 
help determine the parameters in NEI models.
For example, the line ratio between two different elements,
two different ionization states
of the same elements \citep{Vasiliev2011},
or the G-ratio and R-ratio of He-like triplet lines \citep{Vink2012}
can be used. The ratio
between different elements is mainly determined by the abundances.
The G and R-ratio also change with the ionization age. 
In our simulation code, we calculate the 
fractions of different ion states. Therefore, it is easier to 
calculate the line ratio between two ion states, such as
$R=f(\rm{O}^{6+})/f(\rm{O}^{7+})$ 
(See an example in Fig.~\ref{f:discussion} middle left panel).

\begin{figure}[htpb]
  \plottwo{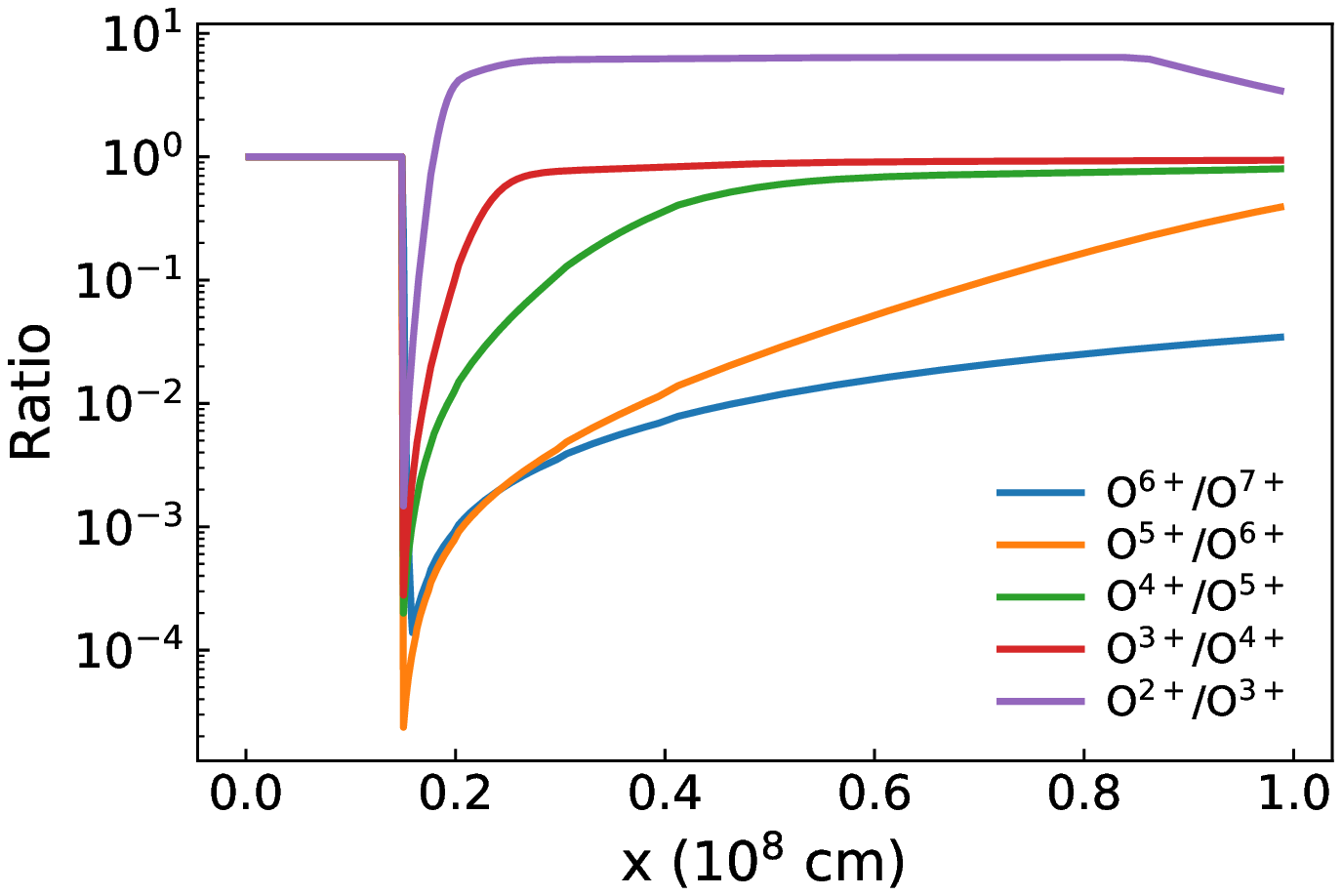}{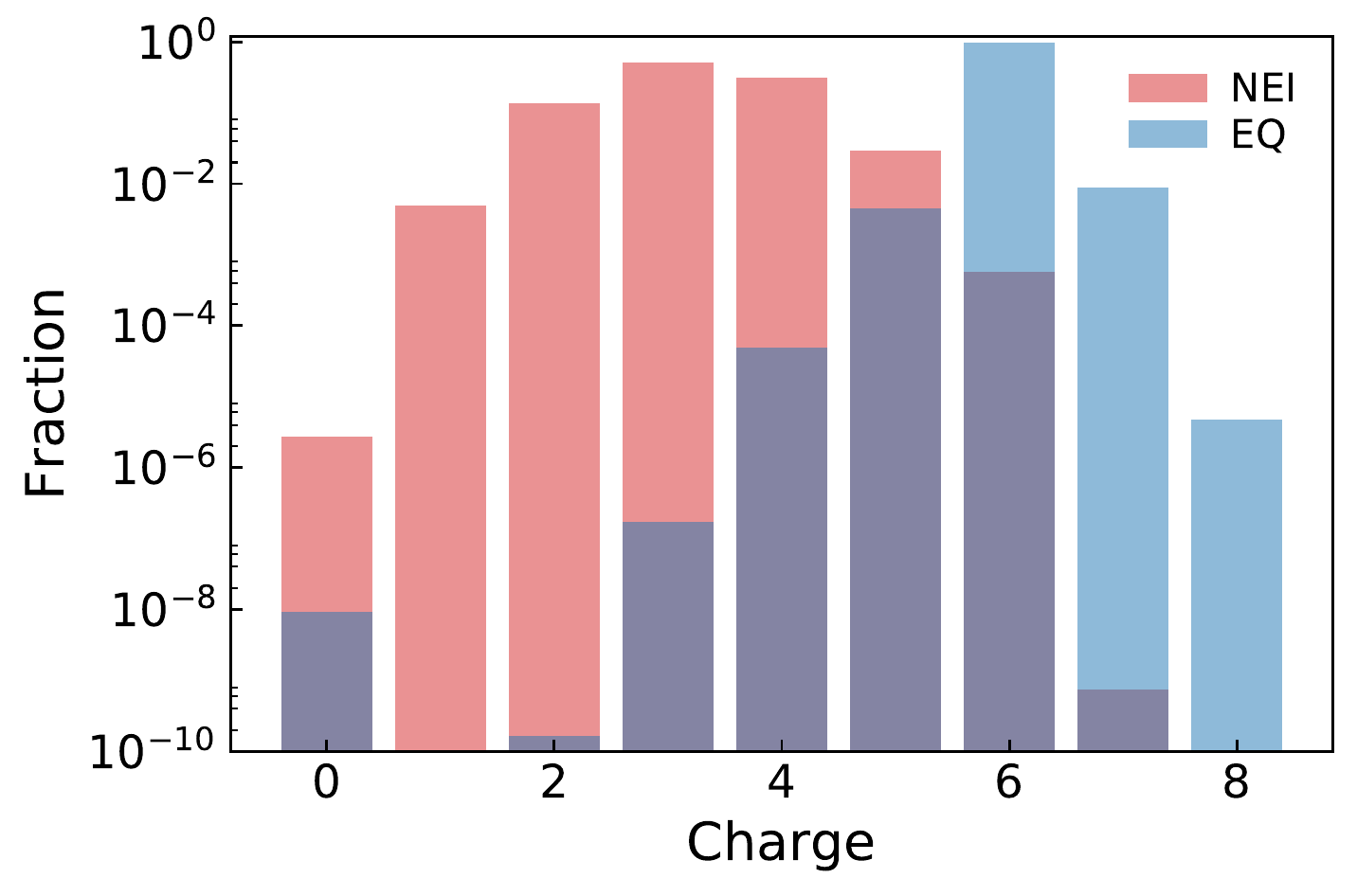}
  \caption{Left: The ratio $r_b=R_{eq}/R$ of different Oxygen ions for {\em NeiTest} simulation. 
  Right: The distribution of oxygen ions in an ionizing state. Purple is the 
  overlapping of red (NEI) and blue (equilibrium).
  \label{f:o_ratios}}
\end{figure}

In a suitable temperature range, the ratio should 
represent the ionization states. 
The fraction ratio in the equilibrium 
to the NEI ratio 
($r_b=R_{eq}/R$; $R_{eq}$ is the same fraction ratio of equilibrium)
can be
considered as a factor for the extent of the ionizing or recombining
status. Assuming that the line fraction ratio can represent the 
temperature, this is a similar strategy used with the ionization balance 
temperature in the SPEX data analysis software\footnote{\url{https://www.sron.nl/astrophysics-spex}}
\citep{Kaastra1996}. A recombining plasma should have $r_b>1$; an ionizing
plasma should have $r_b<1$.
However, the line ratio is not linear with temperature except over a
narrow range, and different
line ratios may show totally different results (See Fig.~\ref{f:o_ratios}).
The ratio between O$^{2+}$ and O$^{3+}$ in Fig.~\ref{f:o_ratios} shows a
``recombining'' feature because the ionization makes the distribution of the ion
fraction move to higher ionization state. The proper ions must be
carefully selected
according to the temperature and ionization state to avoid such a false
conclusion.

\subsubsection{Average charge}\label{sec:averagecharge}
The average charge can be defined by
$C_{ave}=\sum\limits_{i=0}^{Z} f_i c_i$, where
$f_i$ is the ion fraction with the fraction summation normalized to 1, $c_i$
is the charge for the ion, and $Z$ is the atomic number.
It is assumed that ionization increases
the average charge and recombination decreases it. 
Although in some dramatically turbulent plasma, the distribution can be
multimodal (with more than one peak), and the ionizing and recombining
may happen simultaneously, the average charge can still represent the 
overall tendency. \citet{Benjamin2001} even showed it could 
be used to approximate the entire NEI plasma states in certain circumstances.

Similar to the above case, the ratio of average charge between equilibrium
and NEI $r_c=C_{ave,eq}/C_{ave}$
can be used to show the ionization state. The plasma is expected to be
ionizing with $r_c>1$ or recombining with $r_c<1$. However,
in a nearly neutral state, this ratio can be very sensitive to the small 
value in the denominator. In 
Fig.~\ref{f:discussion} middle right panel, some cold clouds are suggested 
to be in a ``strong" recombination while other methods show less 
significance because of the small denominator effect.

Alternatively, the difference between the average charge in
equilibrium and in NEI, 
$d_c=C_{ave,eq}-C_{ave}$, is similar to the
ratio, $r_c$, with the critical value is 0 instead of 1.
When $d_c$ is positive, the plasma is considered to be ionizing; 
when $d_c$ is
negative, the plasma is considered to be recombining.
It can also be written as $d_c=\sum\limits_{i=0}^{Z} (f_{eq,i}-f_i)c_i$,
which is charge-weighted deviation from equilibrium. In this case,
the higher
ionization states (which usually means a higher temperature)
have more influence on this criteria (See 
Fig.~\ref{f:discussion} bottom-left panel for an example). 

\subsubsection{Timescale to achieve equilibrium}\label{sec:timescale}
From the eigenvalue method, the
solutions can be found from Appendix in \citet{Smith2010}
\begin{equation}\label{eq:solution}
{\bm G}=\sum\limits_{i=1}^{n}c_i e^{\lambda_i \tau}{\bm V_R}^{i}
\end{equation}
{ indicating} that the timescale ($n_e t$) can be 
represented by the $e$-folding time as $-1/\lambda_i$ 
\citep[See][for the timescale-temperature relations]{Smith2010}, 
where $\lambda_i$
is the negative eigenvalue for the ion $i$. 
As the density ($n_e$) distribution can also be obtained from the simulation,
the $e$-folding time ($t$) is depicted by $-1/(\lambda_i n_e)$, which is 
dependent on the temperature and density. For a specific element,
the maximum timescale that can be achieved with the maximum $\lambda_i$
implies how much fluence is required to get to equilibrium.
However, with this method, the extent of closeness to equilibrium
is determined
without any information about whether the plasma
is in an ionizing or recombining state.
A combination of the timescale and the other methods are needed 
for showing the ionization states. In the bottom-right panel of 
Fig.~\ref{f:discussion}, this timescale is shown for non-equilibrium 
(the absolute value of average charge difference 
is greater than $1\times 10^{-6}$) regions.
The darker (time is larger) the map is, the further the NEI plasma is
from achieving equilibrium.

\subsubsection{Generating a spectrum}\label{sec:spectrum}
A spectrum with emission from all the ions of interest
can be obtained from simulation results.
This is the most
direct way to compare with observations. With the relevant parameters
for a specific telescope and instrument,
we can mimic an ``observed'' spectrum. By fitting it, the detectable
ionizing or 
recombining features and parameters, e.g.\ timescale, can be directly 
determined. 
But this approach requires
more computing resources than all the previous methods, 
as a complete solution requires a 
3D simulation and the
projection of the 3D simulation. 
This method will be used and shown in a future work.

\bigskip

\begin{figure}[htpb]
  \plottwo{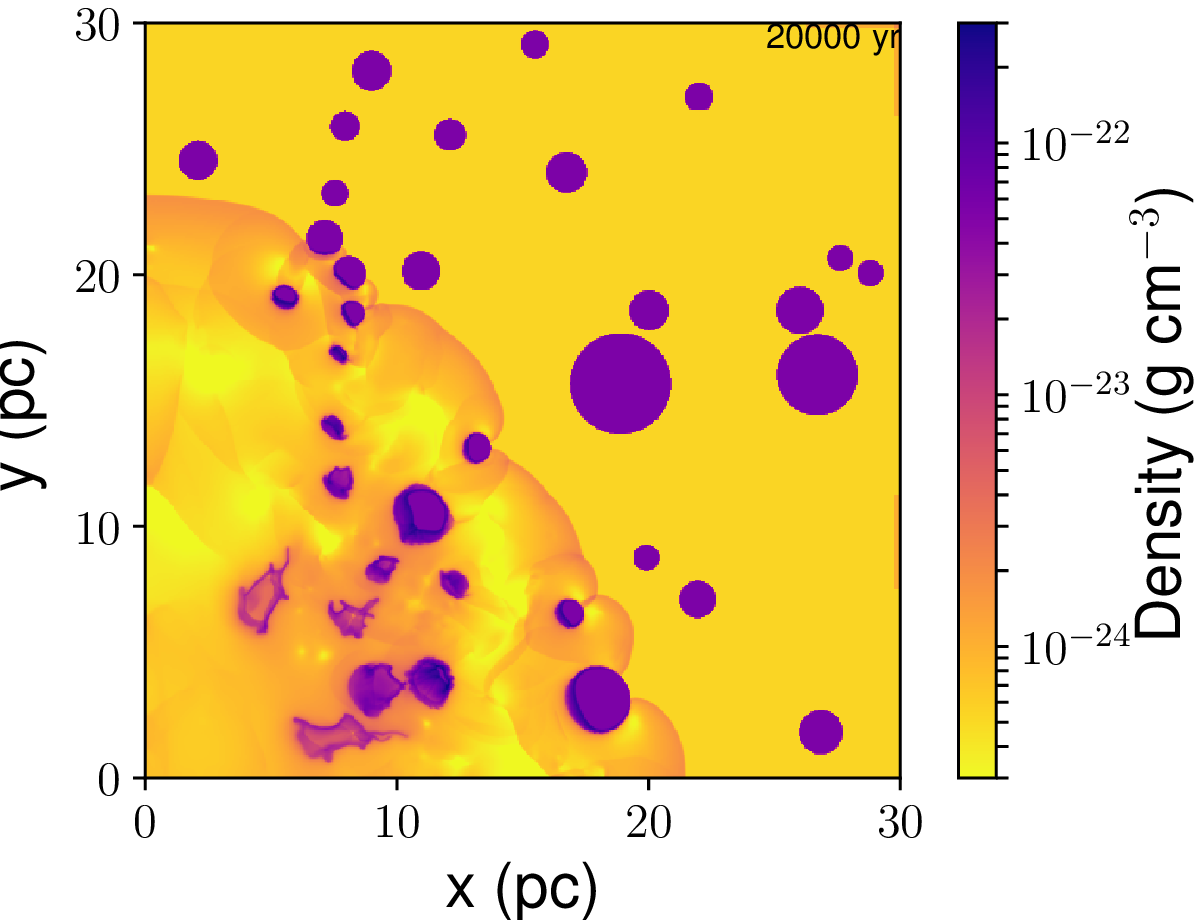}{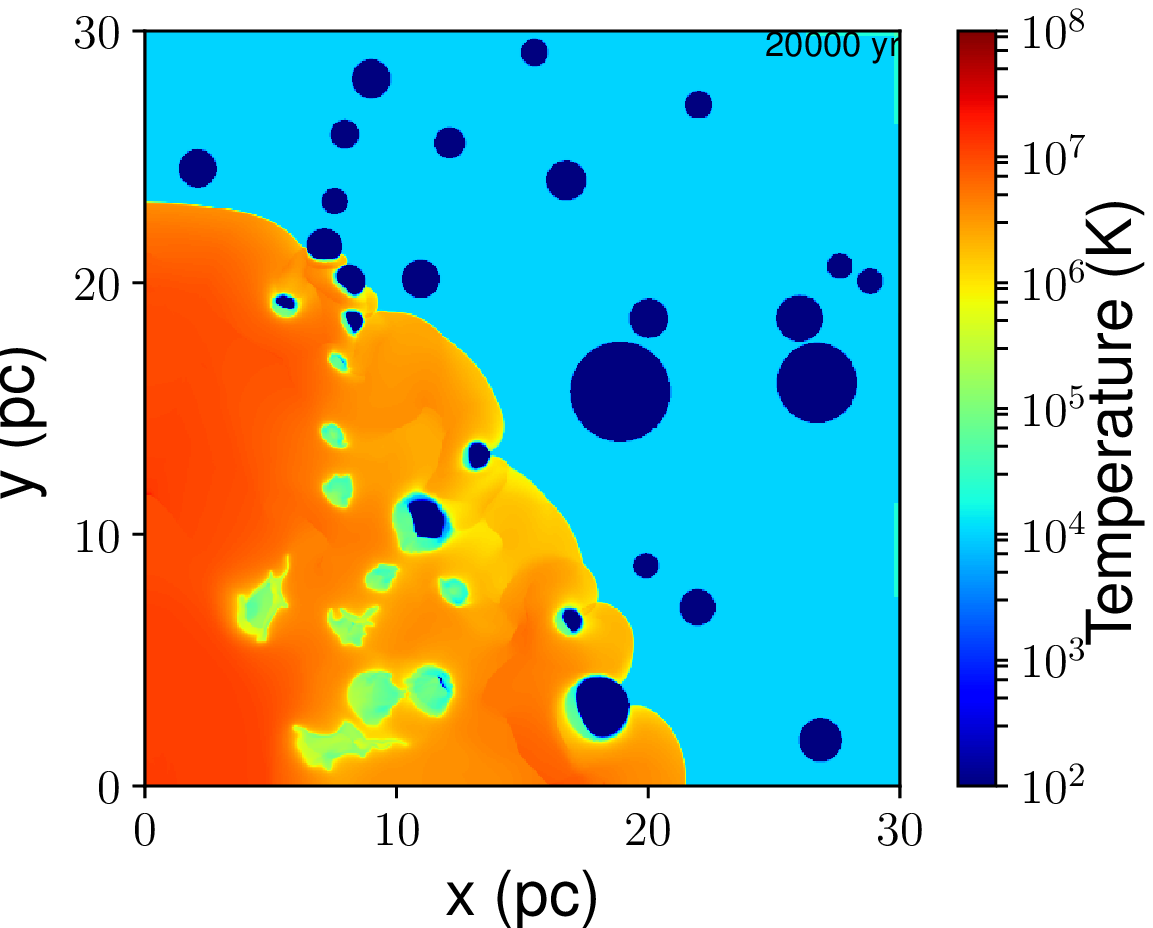}
  \plottwo{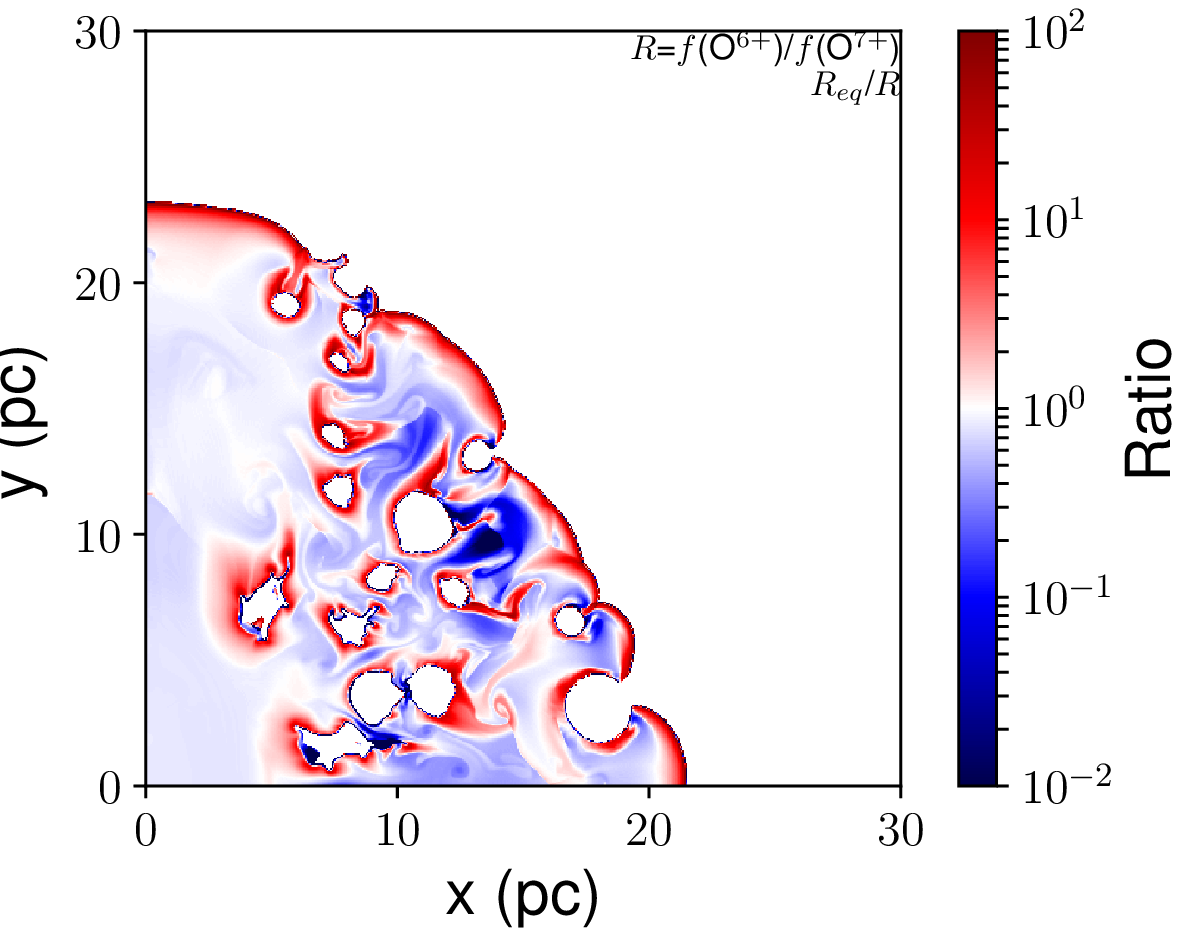}{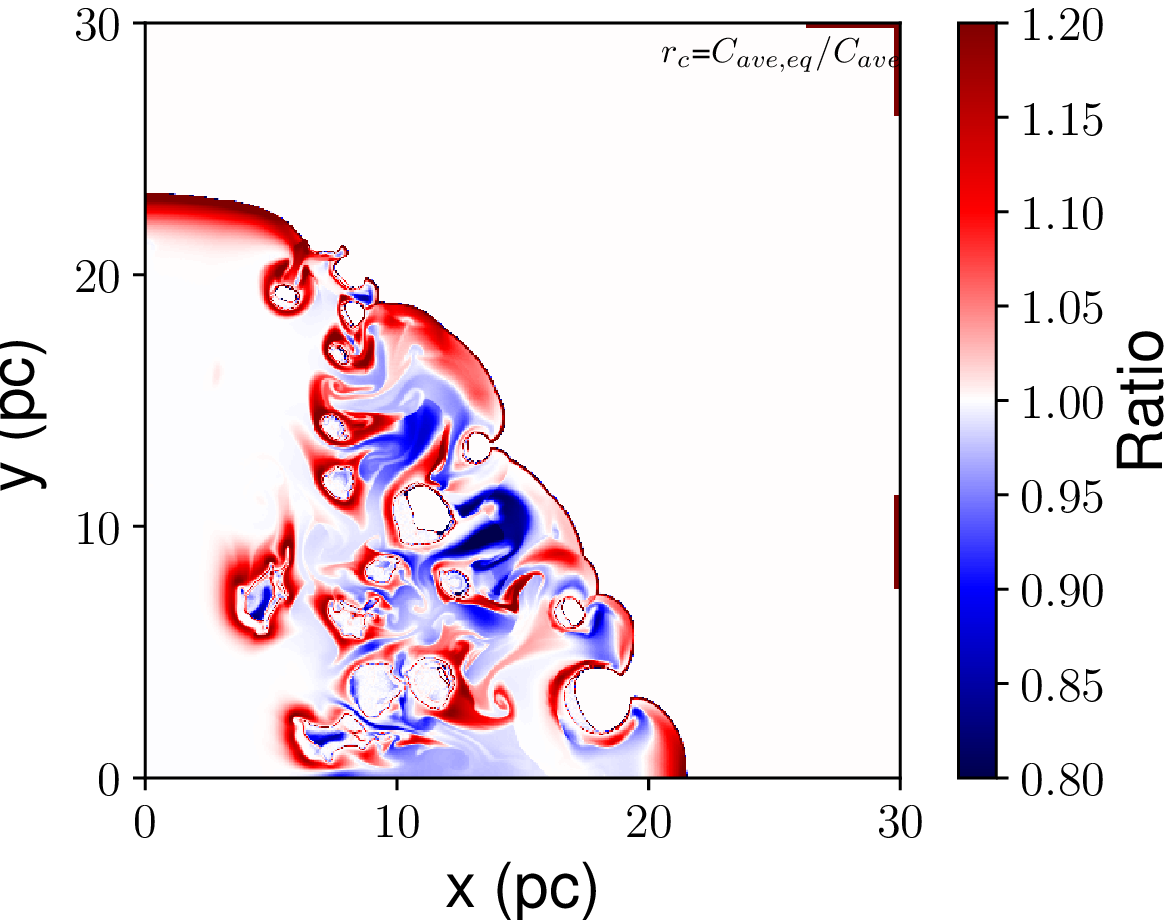}
  \plottwo{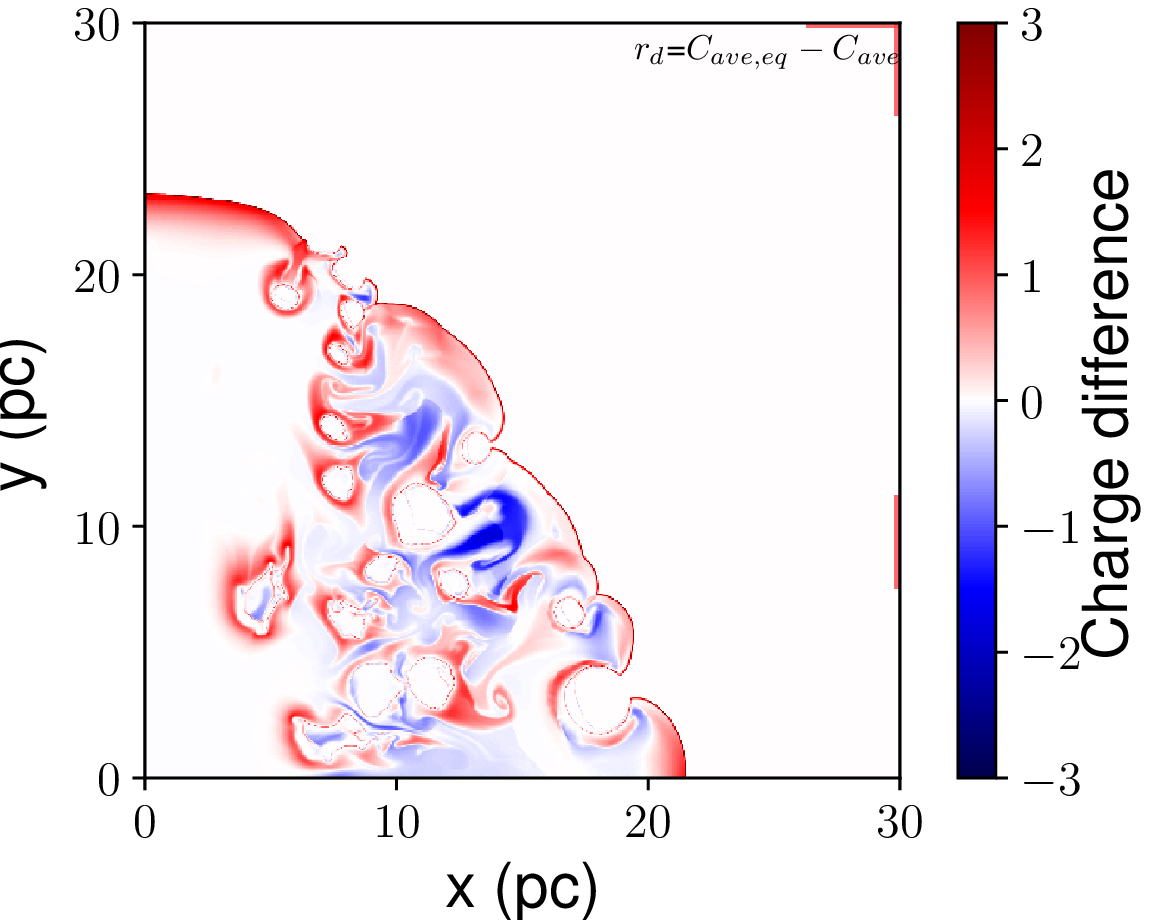}{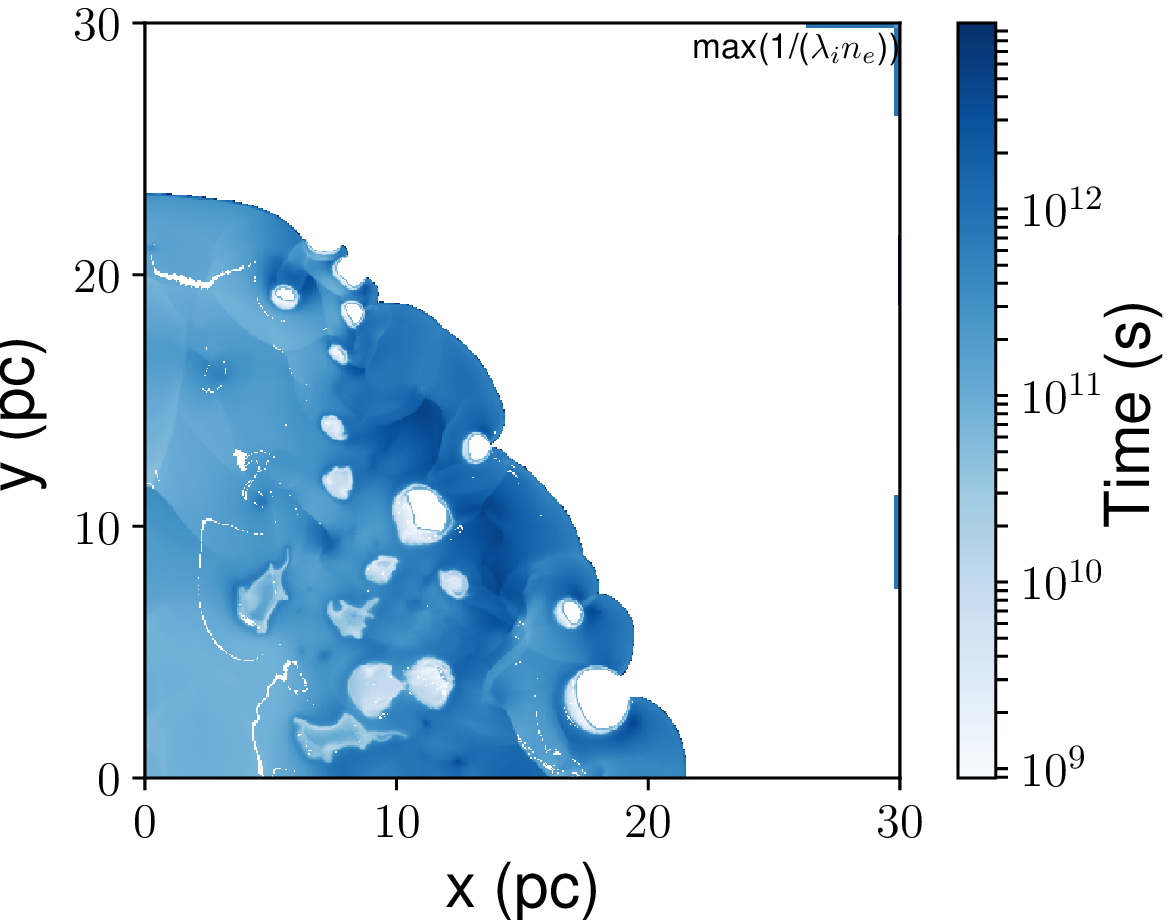}
  \caption{NEI indicators used in an example SNR simulation.
  Top-left: Density distribution; Top-right: Temperature distribution;
  Middle-left: ions ratio (See \S\ref{sec:lineratio}); Middle-right: 
  average charge ratio (See \S\ref{sec:averagecharge}); 
  bottom-left: average charge 
  difference (See \S\ref{sec:averagecharge}); 
  bottom-right: time for $e$-folding ionization/recombination
  (See \S\ref{sec:timescale}).
  \label{f:discussion}}
\end{figure}

\begin{figure}[htpb]
  \plottwo{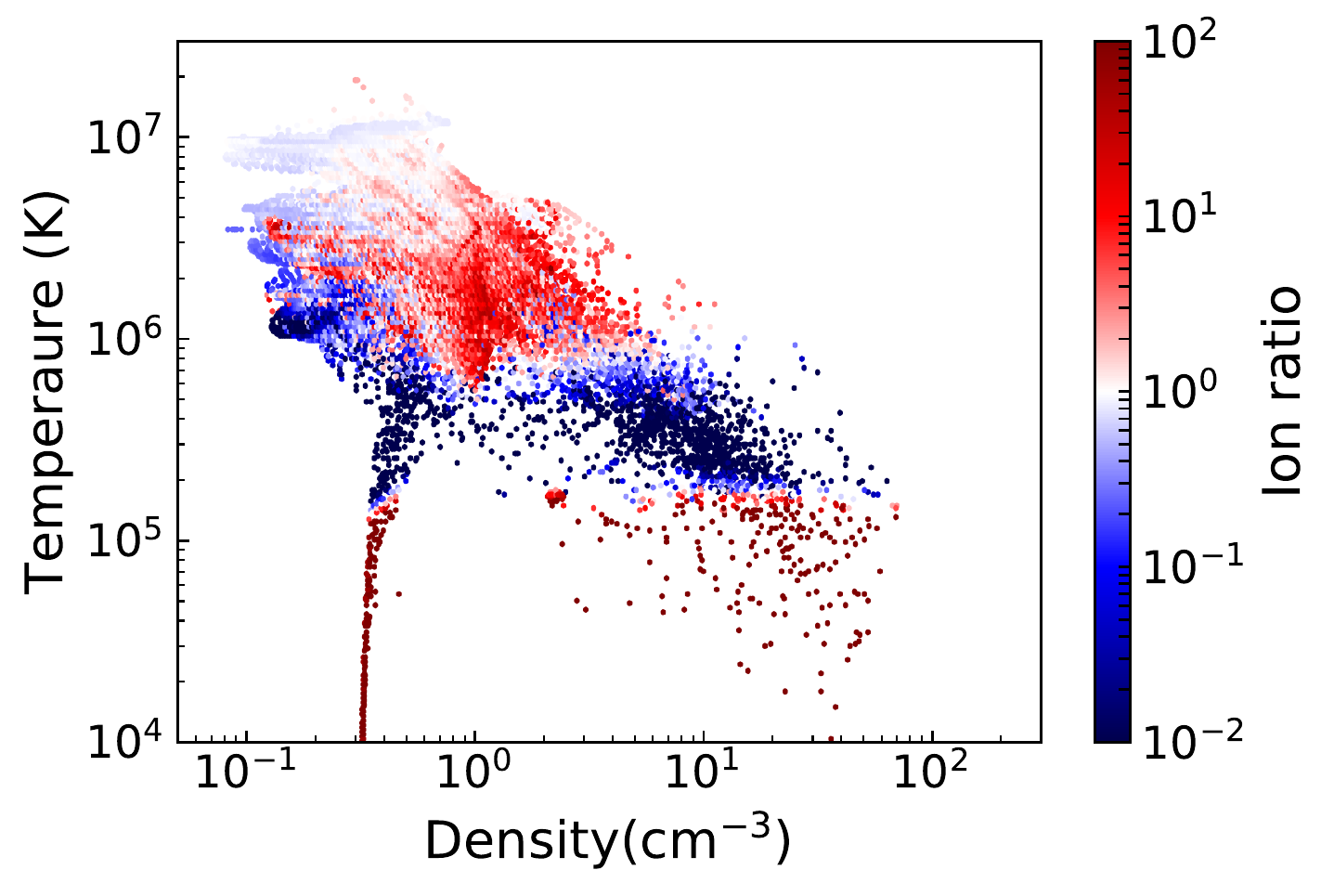}{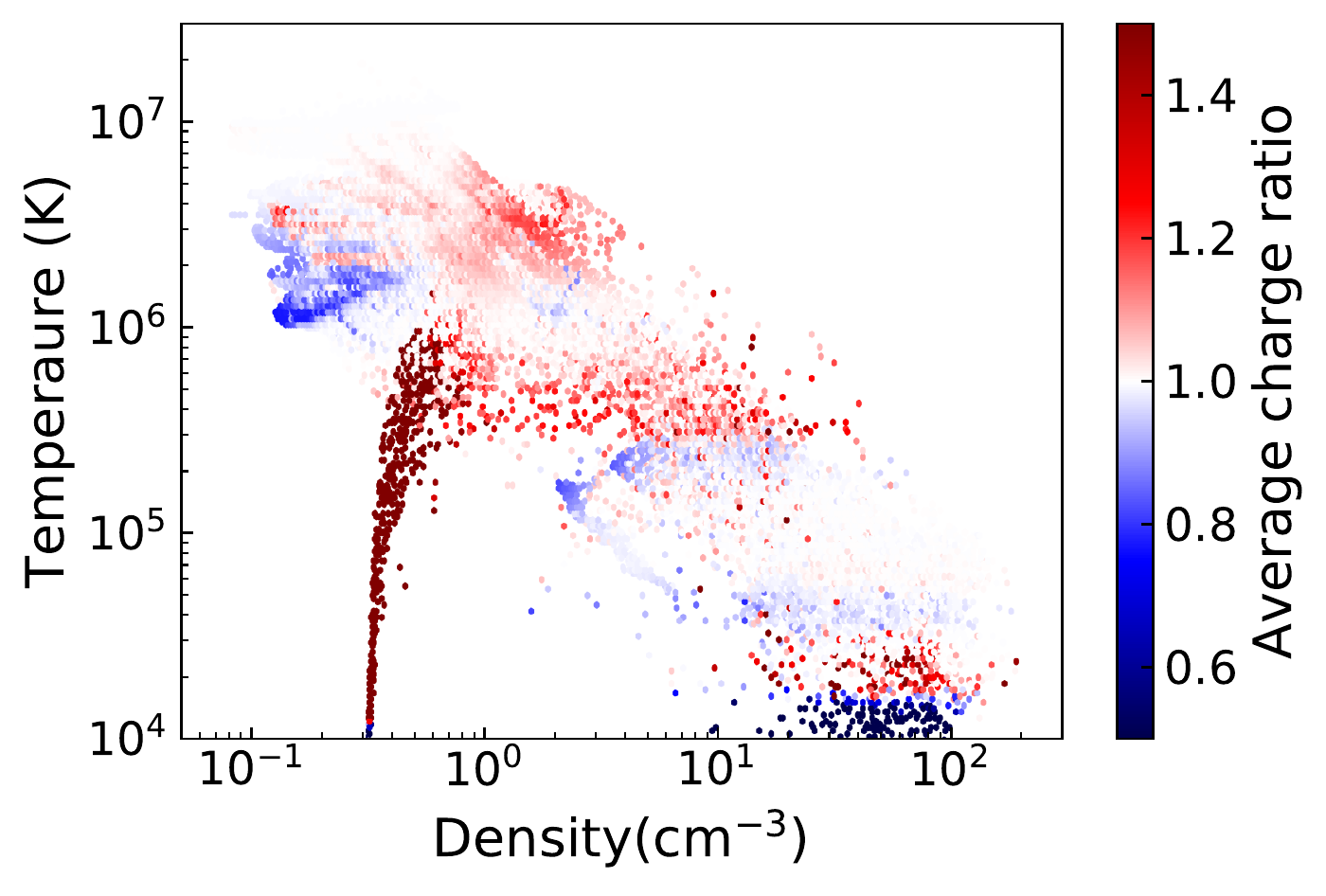}
  \plottwo{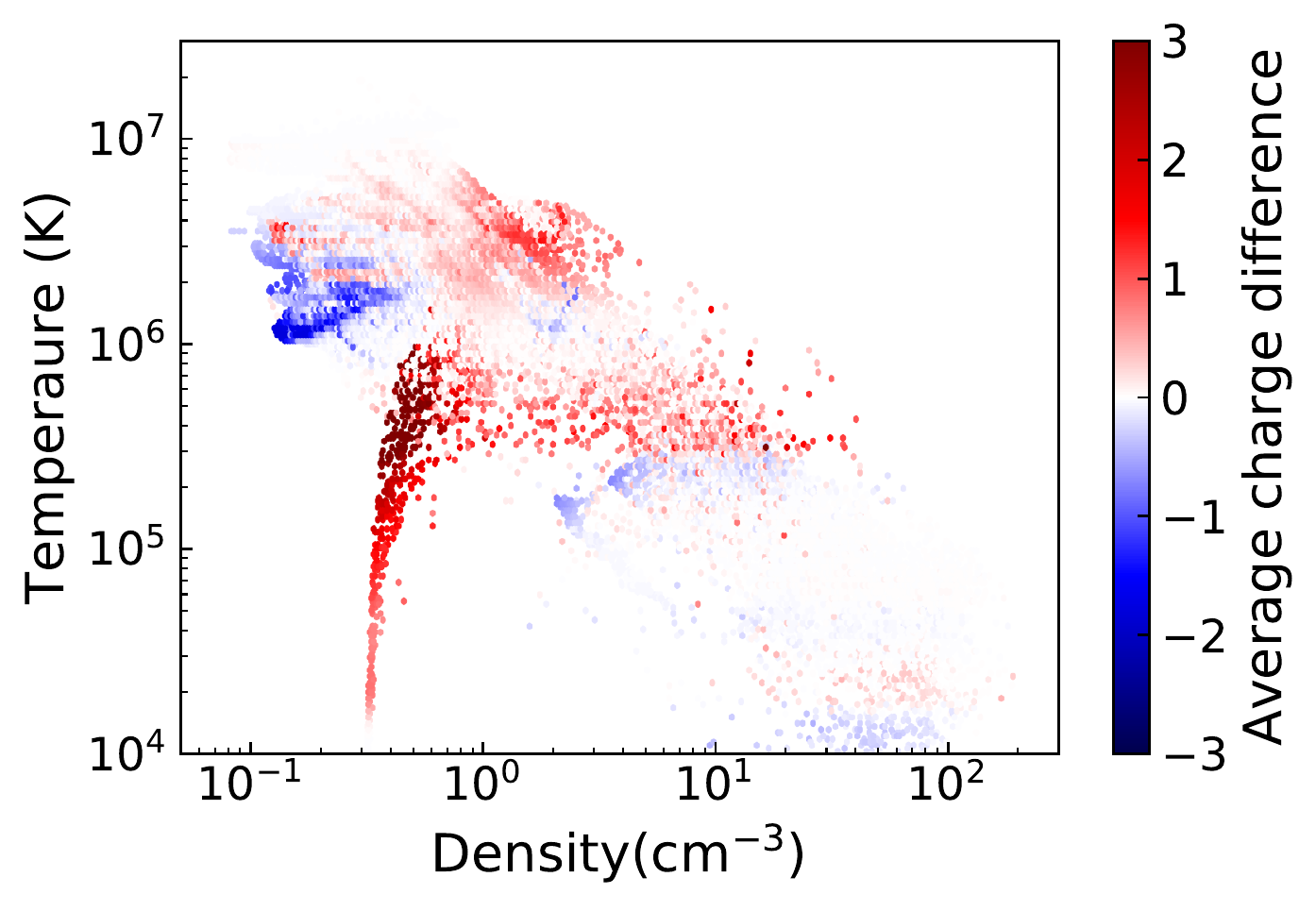}{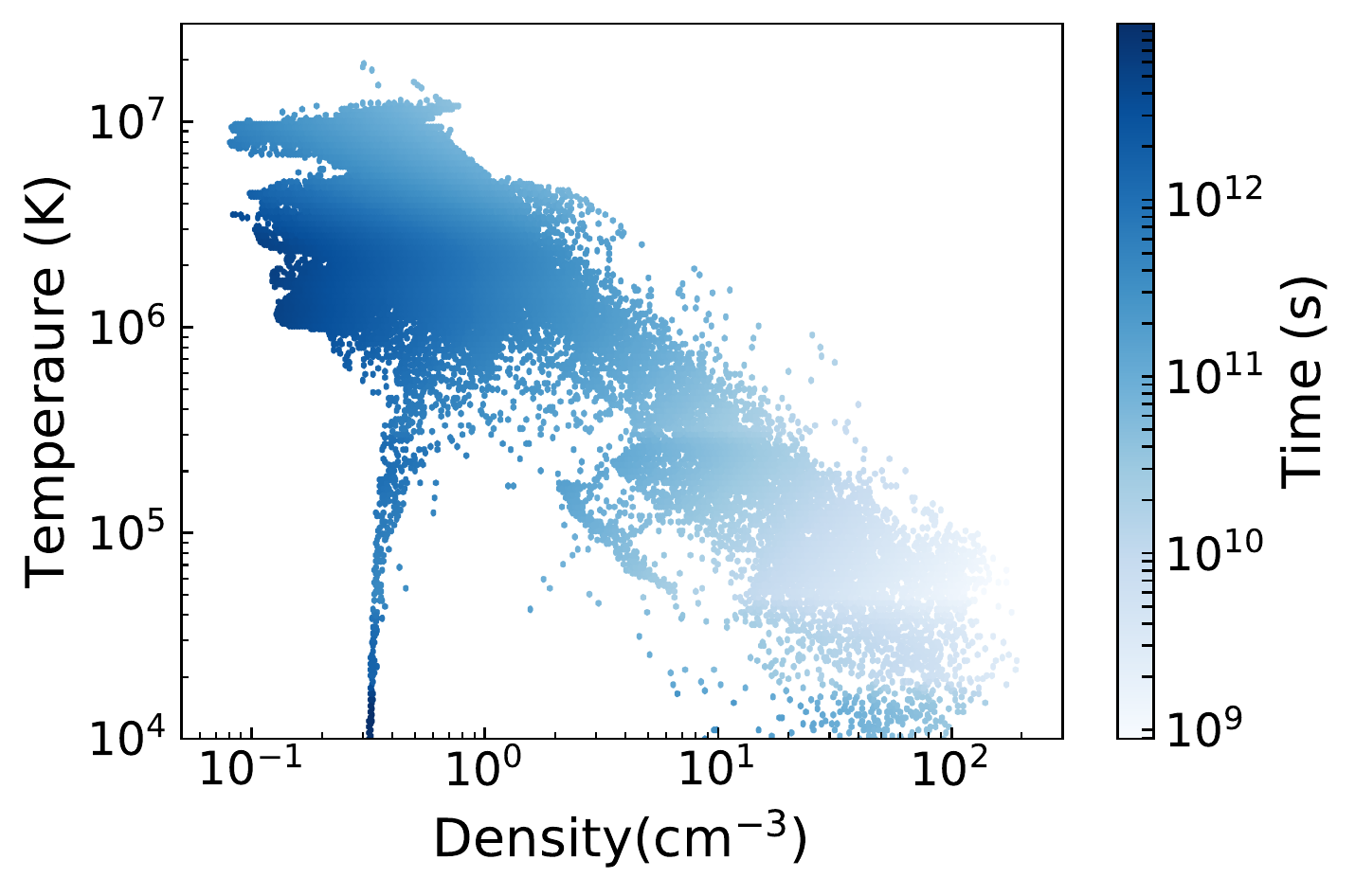}

  \caption{Density-temperature phase diagrams of the example simulation shown 
  in Fig.~\ref{f:discussion}. The four panels are 
  correspondent to the lower four 
  panels in Fig.~\ref{f:discussion} one by one, that is (from left to right,
  top to bottom) ions ratio of O$^{6+}$ and O$^{7+}$, average charge ratio,
  average charge difference, and the time of $e$-folding 
  ionization/recombination. 
  \label{f:discussion2}}
\end{figure}

Except generating a spectrum,
we recommend the average charge difference for the determination of
an NEI state. 
In Fig.~\ref{f:discussion2}, we plot the density-temperature phase 
diagrams for the methods we mentioned above. All the pixels in the example
simulation (Fig.~\ref{f:discussion}) are
shown in a scatter plot in the density-temperature
plane that shows the average density and temperature in each MHD
``pixel''. 
First, 
we can see that ion ratio is effective only for a special temperature range.
Below about 6$\times 10^5\K$, the ratio of O$^{6+}$ and O$^{7+}$ is only
a function of temperature. The ``recombining'' points in the lower temperature 
range are misleading. Some of them are in the shock front itself,
where the temperature and ionization state are low but rapidly rising.
In the top-right panel of 
Fig.~\ref{f:discussion2}, the average charge ratio is significant when 
the temperature is low and the average charges for both the ions state and the
equilibrium are near to zero. To exclude the possibility of a 
zero-divided value,
the average charge difference method is recommended for the depiction
of ionization state. Sometimes only 
the ionizing or recombining states are important 
while how far it is from the equilibrium is not. Both the average charge ratio
and average charge difference can satisfy the requirements with
different critical values (1 and 0 respectively). 

\subsection{NEI in Eulerian code}
The FLASH code is primarily an Eulerian code. 
The NEI solver is separated from the FLASH hydrodynamic solver to 
get rid of the advection term in Eulerian fluid equation (See FLASH
code 
manual\footnote{\url{http://flash.uchicago.edu/site/flashcode/user_support/flash4_ug_4p3.pdf}} \S~16.2.1 for details). The ion fractions are firstly 
transported spatially without the ``source''
term (ionizing and recombining). After each transport step,
an ordinary differential equation is solved in each cell with the 
elapsed time step, which is a first order approximation.
In this version of the FLASH code (FLASH4), the particles in Lagrangian 
scheme can be used to trace the flow.
The NEI code may be also used on the particles in a
future work to see how it differs from the Eulerian code.

\subsection{Applications to other MHD codes}

Currently, the NEI calculation code is written in free-format
Fortran code, compatible with
the FLASH codes. For other MHD simulation code 
(e.\ g., ZEUS, ATHENA), the NEI calculation code can be compiled 
to a Fortran module to integrate with the MHD codes.
The atomic data is stored in FITS files \citep{Wells1981}, and is available from AtomDB
as a standard product at \url{http://atomdb.org/download.php}.

\section{Summary}

In this paper, we have described an optimized method, the eigenvalue method,
to calculate NEI
in an MHD simulation (FLASH code) with an updated atomic database.  
The updated database is compared with 
the original to 
show that the differences are large and the update is necessary. 
The new eigenvalue method is also compared to the original one by using
the same updated database to show that the results
are consistent to each other and the efficiency can be greatly improved.
Radiative cooling of the ions used in the simulation can be included 
with the eigenvalue method in the simulation. We discussed the ways to
measure the ionization states. With an example simulation, the average 
charge difference is shown as a better method. 

\section*{Acknowledgement}
G.Y.Z.\ is grateful to the support of the CSC, 
the 973 Program grants 2017YFA0402600 and 2015CB857100,
and NSFC grants 11773014, 11633007, 11233001 and 11503008.

\software{FLASH code \citep{Fryxell2000}, yt \citep{Turk2011}}

\bibliographystyle{aasjournal}
\bibliography{cite}

\end{CJK*}
\end{document}